\RequirePackage{fix-cm}
\documentclass[12pt]{article}
\usepackage{arxiv}
\usepackage{graphicx}
\usepackage{epsf}
\usepackage{subfig}
\usepackage[numbers]{natbib}
\usepackage{color}
\usepackage[colorlinks=true,linkcolor=blue]{hyperref}
\usepackage{url}
\usepackage[T1]{fontenc}
\usepackage[normalem]{ulem}
\usepackage{cancel}
\usepackage{mathrsfs}
\usepackage{bbm}
\usepackage{stmaryrd}
\usepackage{bm}
\usepackage{amsfonts}
\usepackage{amssymb}
\usepackage{amsmath}

\newcommand{\gaeq}{\raisebox{-.7ex}{$\stackrel{\textstyle>}{\sim}$}\ }
\newcommand{\exgra}{{\em e.g.}}
\newcommand{\idest}{{\em i.e.}}

\begin{document}

\title{Nonlinear dynamics and phase space transport by chorus emission}
%\subtitle{Nonlinear dynamics and phase space transport by chorus emission}

% \titlerunning{Chorus emission and phase space structures}  % if too long for running head

\author{
  Fulvio Zonca\thanks{\texttt{fulvio.zonca@enea.it}}\\
  Center for Nonlinear Plasma Science and C.R. ENEA Frascati, CP 65-00044 Frascati, Italy
  \And
  Xin Tao\thanks{\texttt{xtao@ustc.edu.cn}}\\
  Department of Geophysics and Planetary Sciences, University of Science and Technology of China, Hefei, China \\
  \And
  Liu Chen\thanks{\texttt{liuchen@zju.edu.cn}} \\
  Institute of Fusion Theory and Simulation, Department of Physics, Zhejiang University, Hangzhou, China
}

% \date{Received: \today / Accepted: date }

\maketitle

\begin{abstract}

Chorus emission in planetary magnetospheres is taken as working paradigm to motivate a short
tutorial trip through theoretical plasma physics methods and their applications.
Starting from basic linear theory, readers are first made comfortable with whistler wave packets
and their propagation in slowly varying weakly nonuniform media, such as the Earth's magnetosphere,
where they can be amplified by a population of supra-thermal electrons. The nonlinear
dynamic description of energetic electrons in the phase space in the presence of
self-consistently evolving whistler fluctuation spectrum is progressively introduced 
by addressing renormalization of the electron response and spectrum evolution
equations. Analytical and numerical results on chorus frequency chirping are obtained and 
compared with existing observations and particle in cell simulations. Finally, the general 
theoretical framework constructed during this short trip through chorus physics is used
to draw analogies with condensed matter and laser physics as well as magnetic confinement
fusion research. Discussing these analogies ultimately presents
plasma physics as an exciting cross-disciplinary field to study.

%***252  mots but: < 250 mots
% \keywords{%First keyword \and Second keyword \and More
% Whistler waves \and Nonlinear phenomena \and Classical field theories}
% \PACS{52.35.Hr \and 52.35.Mw \and 03.50.Kk}
\end{abstract}

\section{Introduction}
\label{sec:introduction}

Chorus emission, observed in various planetary magnetospheres \cite{Tsurutani1974,Burtis1976, Hospodarsky2008,Menietti2008c}, is one of the
long standing and exciting problems of space plasma physics, which is still attracting significant interest because of the many implications it bears
as well as of its practical applications. One of the striking features of chorus is its close analogy with neighboring
areas of plasma and condensed matter physics, such as laser research \cite{SotoChavez2012} and 
magnetic confinement fusion \cite{Zonca2015b,Zonca2015,Chen2016}. 
The common roots of all these phenomena are nonlinear dynamics and 
phase space transport, and using well-established methods of classical field theory is the key to 
emphasize their general aspects \cite{zonca18}. Sharing these views with a broad readership is the main
motivation of this work. Thus, the present scope is not writing a specifically focused review paper; 
but rather a tutorial work on the general foundations of plasma theory with chorus emission as a paradigmatic application. 
That  is, in this work, we will illustrate how classical field theoretical methods can yield a comprehensive theoretical framework 
for discussing nonlinear dynamics as well as phase space transport; and, finally,  as an example of practical application, 
providing in-depth analyses of chorus emission and frequency chirping. 
Exploring the additional complexities
due to magnetic field configurations and plasma equilibrium non-uniformities, \exgra, in magnetic
fusion plasmas, would then come naturally as a further step, where technical complications would
not be at risk of obscuring the underlying physics.

The structure of this work is intended as a short trip, accompanying readers from the linear theory of 
whistler wave excitation by supra-thermal electrons in Sec. \ref{sec:choruslin}, through the theoretical formulation
of phase space transport and nonlinear dynamics in Sec. \ref{sec:phasespace}, and finally arriving
at the application to chorus emission and frequency chirping in Sec. \ref{sec:chorus}. 
The material is presented to be accessible to the broadest readership, offering different levels of
articulation and depths that should leave readers free to choose their own favorite combination. The linear theory
of Sec. \ref{sec:choruslin} touches upon the linear dispersion relation of  whistler waves
constructed from the usual cold plasma dielectric tensor (cf. Sec. \ref{sec:whistler}). Then, 
it addresses the resonant destabilization by a ``hot'' or ``supra-thermal'' electron component (cf. Sec.
\ref{sec:whistlerinst}) and, finally, it analyzes the whistler wave packet propagation in a slowly
varying weakly nonuniform medium (cf. Sec. \ref{sec:whistlerwp}).  As it is often the case,
linear theory allows us to draw strong conclusions of general validity, such as chorus chirping
being necessarily a nonlinear process. The more advanced topics of
Sec. \ref{sec:phasespace} are gradually introduced starting, in Sec. \ref{sec:singlep}, from the analysis of single particle
motion in a finite amplitude whistler wave packet. Many of the chorus qualitative features and their connection
with phase space structure formation (holes and clumps) and frequency chirping are discussed 
already at this level. More advanced analysis of phase space nonlinear dynamics are presented
in Sec. \ref{sec:smallamp}, where we introduce the solution of kinetic equations based on small 
fluctuation amplitude expansion \cite{vanhove55,prigogine62,balescu63,Altshul1966} and derive a Dyson-like equation \cite{Dyson1949,Schwinger1951} 
for the renormalized particle response \cite{Zonca2017,Zonca2021},
as well as the well-known resonance broadening theory \cite{Dupree1966,aamodt67,weinstock69,Mima73},
based on Dupree's seminal work on perturbation theory for strong plasma turbulence \cite{Dupree1966}. Here, some readers may be satisfied to qualitatively
follow the physics concepts and theoretical framework. Some others, may find
all the necessary details to dive deeper into the foundations of this approach. General conservation properties
of phase space transport equations are then given in Sec. \ref{sec:conserve}, along with 
a qualitative discussion of the variety of dynamic behaviors that are expected for the whistler fluctuation
spectrum excited by supra-thermal electrons. This discussion actually anticipates some of the 
contents on chorus emission and frequency chirping of a brief presentation on chorus 
observation and qualitative features given in Sec. \ref{sec:chorusobs}
as opening of Section \ref{sec:chorus}.
For the sake of simplicity, and as illustration of the strength of the present approach, Sec. \ref{sec:reddyson_der}
gives a reduced description of the general theoretical framework introduced in Sec. \ref{sec:phasespace}, which
yields nonlinear evolution equations for the driving rate of the fluctuation spectrum and corresponding nonlinear frequency shift
induced by resonant supra-thermal electrons. Notably, these equations allow us to analytically derive the
expression of chorus chirping rate originally conjectured by \cite{Vomvoridis1982} on the basis of 
simulation results. Section \ref{sec:reddyson_sol} then provides a numerical solution of the 
aforementioned reduced description, showing qualitative and quantitative agreement with simulation 
results by the DAWN code \cite{Tao2017a} with the same parameters. A comparison of the understanding
of chorus chirping within the present theoretical framework with other existing models, \exgra \ 
\cite{Helliwell1967,Sudan1971,Vomvoridis1982,Omura2011,Tsurutani2020,Wu2020,Tao2021}, is also presented
in Sec. \ref{sec:compare}. Finally, Sec. \ref{sec:conclusions}, gives a comparative discussion of 
analogies of chorus emission and phase space nonlinear dynamics with super-radiance in free electron
lasers \cite{bonifacio90,bonifacio94,giannessi05,watanabe07} 
and current research topics in magnetic confinement fusion \cite{Zonca2015b,Zonca2015,Chen2016} 
along with concluding remarks.

\section{Linear excitation of whistler waves by supra-thermal electrons}
\label{sec:choruslin}

Whistler waves are very low frequency R-X mode electromagnetic waves
[e.g., \cite{Stix1992}, page 37] with frequencies between lower hybrid
resonance frequency and electron cyclotron frequency, which, in the
Earth's magnetosphere, is about a few kHz to tens of kHz. Their name
was motivated by lightning generated whistlers, which, due to
dispersion, were typically perceived as a descending tone that could
last a few seconds. Typical naturally occurring whistler mode waves
include lightning whistlers, plasmaspheric hiss, lion roars and
chorus. The focus of the present brief tutorial, chorus emission, is
generated by a nonlinear process as will be shown in the following.
Whistler waves can propagate parallel as well as obliquely with respect to the Earth's magnetic field, although 
oblique whistler waves are characterized by smaller wave magnetic field amplitudes but larger electric field amplitudes than parallel waves 
\cite{Artemyev2016b}.

Here, we first introduce the fundamental linear properties of transverse whistler waves 
that propagate parallel to the Earth's magnetic field. Invoking the separation of scales
between wavelength and equilibrium non-uniformity scale length, we adopt the representation
of fluctuations as wave packets and derive the linear dispersion relation in Sec. \ref{sec:whistler}.
Section \ref{sec:whistlerinst} then discusses the whistler wave excitation mechanism by a
population of supra-thermal electrons characterized by velocity-space anisotropy.
In particular, Eq. (\ref{eq:WGamma4lin}) summarizes the expressions for wave packet growth and
frequency shift, which also hold nonlinearly, as shown in Sec. \ref{sec:chorus}.
We finally show, in Sec. \ref{sec:whistlerwp}, that whistler wave packet intensity and 
phase evolution equations, Eqs. (\ref{eq:actionevolve}) and (\ref{eq:phaseevolve}),
can be integrated exactly by means of the method of characteristics. These results 
illustrate the linear dispersive properties of whistler waves and the spatiotemporal
features of their linear propagation; \exgra, showing that whistler wave packets 
excited by supra-thermal electrons are a convective instability. Further to this,
they allow us to conclude that chorus frequency chirping can be clearly identified 
as a nonlinear process, thereby motivating the further analysis that will be presented
in later sections. 

\subsection{Whistler wave dispersion relation}
\label{sec:whistler}

Let us consider a transverse whistler wave propagating parallel to the Earth's magnetic field. Due to the large separation
between wavelength and magnetic field line scale length, we can describe whistler waves as wave packets
propagating in a weakly nonuniform medium \cite{Bernstein75,Kravtsov90,Stix1992}, where
\begin{equation}
\delta \bm E_\perp (z,t)  = \frac{1}{2} \sum_k \left(  e^{i S_k(z,t)} \delta \bar{\bm E}_{\perp k}(z,t) + c.c. \right) \; , 
\label{eq:choruswp}
\end{equation}
$\perp$ denotes the component transverse to the ambient Earth's magnetic field, $z$ is the coordinate along the magnetic field; and the eikonal $S_k$ is related with the (parallel) wave number $k = \partial_z S_k$ and frequency $\omega = - \partial_t S_k$ of the wave packet itself. 
From Faraday's law, we have, assuming parallel propagation of transverse wave,
$$\delta \bar{\bm B}_{\perp k}= \frac{k c}{\omega} \hat{\bm z} \times \delta \bar{\bm E}_{\perp k}(z,t) \; ; $$
with $\hat{\bm z}$ the unit vector along the field line, $\delta \bar B_{y k} = (k c/\omega)\delta \bar E_{x k}$ and
$\delta \bar B_{x k} = - (k c/\omega)\delta \bar E_{y k}$. Meanwhile, from Amp\`ere's law, and separating the current 
density perturbation $\delta \bm J$ in contributions carried by ``core'' (c) and ``hot'' (h) electron components, 
we have 
\begin{equation}
\left( 1 - \frac{k^2 c^2}{\omega^2} \right) \delta \bar{\bm E}_{\perp k}  + \frac{4\pi i}{\omega} \delta \bar{\bm J}_{c k} \equiv \bm \epsilon_\perp \cdot \delta \bar{\bm E}_{\perp k}- \frac{k^2 c^2}{\omega^2} \delta \bar{\bm E}_{\perp k} = - \frac{4\pi i}{\omega} \delta \bar{\bm J}_{h k}\; , \label{eq:amperemod0}
\end{equation}
where, by analogy with Eq. (\ref{eq:choruswp}), we let
\begin{equation}
\delta \bm J_h (z,t)  = \frac{1}{2} \sum_k \left(  e^{i S_k(z,t)} \delta \bar{\bm J}_{h k}(z,t) + c.c. \right) \; ;
\label{eq:Jhwp}
\end{equation}
and similar representation was assumed for the core electrons current $\delta \bm J_c (z,t)$.
Considering ``core'' ions as a neutralizing background carrying negligible current, and 
``core'' electrons as a cold fluid with density $n$, $\bm \epsilon_\perp$ becomes the 
usual cold plasma dielectric tensor 
\begin{equation}
\bm \epsilon_\perp \cdot \delta \bar{ \bm E}_{\perp k} = \left( 1 + \frac{\omega_p^2}{\Omega^2 - \omega^2}\right) \delta \bar{ \bm E}_{\perp k}
- i \frac{\Omega}{\omega} \frac{\omega_p^2}{\Omega^2 - \omega^2} \hat {\bm z} \times \delta \bar{ \bm E}_{\perp k} \; , \label{eq:coldplasma}
\end{equation}
with $\omega_p^2 = 4\pi ne^2/m$ the electron plasma frequency 
and $\Omega = e B/(mc)$ the electron cyclotron frequency, having denoted with $e$ the positive electron 
charge and with $m$ the electron mass. Assuming a whistler wave packet with right circular polarization,
that is with $\delta \bm E_{\perp}$ co-rotating with the electron cyclotron motion,
$\delta \bar E_{y k} = i \delta \bar E_{x k}$ and $\delta \bar B_{y k} = i \delta \bar B_{x k}$. This yields:
\begin{equation}
\bm \epsilon_\perp \cdot \delta \bar{\bm E}_{\perp k}\simeq \left( 1 + \frac{\omega_p^2}{\omega(\Omega - \omega)}\right) \delta \bar{\bm E}_{\perp k} \; . \label{eq:coldchorus}
\end{equation}
Thus, by direct substitution of this latter expression into Eq. (\ref{eq:amperemod0}), the problem of transverse whistler wave packet with right circular polarization and interacting with hot electrons can be approximately cast as (cf., \exgra, \cite{Nunn1974,Omura1982,Omura2011,Omura2008})
\begin{equation}
\left( 1 + \frac{\omega_p^2}{\omega(\Omega - \omega)}\right) \delta \bar{\bm E}_{\perp k}  - \frac{k^2 c^2}{\omega^2} \delta \bar{\bm E}_{\perp k} = - \frac{4\pi i}{\omega} \delta \bar{\bm J}_{h k}
\; , \label{eq:master} \end{equation}
where, due to the low density of supra-thermal electrons, the right hand side can be formally treated as a perturbation to the lowest order propagation of the 
whistler wave packet. Equation (\ref{eq:master}) allows us to introduce the whistler wave dielectric constant, $\epsilon_w$, and dispersion function, 
$D_w$, such as 
\begin{equation}
\epsilon_w = 1 + \frac{\omega_p^2}{\omega(\Omega - \omega)} \; , \;\;\;\;\; D_w = \epsilon_w  - \frac{k^2 c^2}{\omega^2} \; .  \label{eq:chorusdef}
\end{equation}
Consistent with Eq. (\ref{eq:master}), whistler waves excited by supra-thermal electrons satisfy the lowest order 
WKB dispersion relation
\begin{equation}
D_w(z,K(z,\omega),\omega) = 0 \; , \label{eq:chorusdisp}
\end{equation}
where, by $K(z,\omega)$ we have denoted the wave number solution of Eq. (\ref{eq:chorusdisp}) as a function of $z$ parameterized by $\omega$.
It can be shown that the whistler wave group velocity is given by
\begin{equation}
v_g = \frac{\partial \omega}{\partial k}  =  \frac{2 (\Omega - \omega)^2 k c^2}{ 2 (\Omega - \omega)^2 \omega + \omega_p^2 \Omega} \; ; \label{eq:vg}
\end{equation}
thus, phase and group velocities have the same sign. Meanwhile, for $\omega^2 \sim \Omega^2 \ll \omega_p^2$, it can
be verified that $0\leq \omega < \Omega$.

The excitation mechanism by an anisotropic distribution of ``hot'' electrons is briefly discussed in the next subsection.

\subsection{Whistler instability via wave particle resonant interactions}
\label{sec:whistlerinst}

As anticipated in the previous section, Eq. (\ref{eq:master}) describes the dynamics of a  transverse whistler wave packet with right circular polarization  
interacting with hot electrons. In particular, it is the ``hot'' electron perpendicular current on the right hand side that allows energy transfer from the 
supra-thermal electrons to the whistler wave packet, thereby causing the instability via wave particle resonant interactions. It is readily verified by inspection
that the wave-particle power transfer is controlled by $\delta \bar{\bm J}_{h k} \cdot \delta \bar{\bm E}_{\perp k}^*$.

In order to calculate the hot electron perpendicular current, let us consider the linearized Vlasov equation for the electron distribution function
$f = f_0 + \delta f$ 
\begin{equation}
\left( \frac{\partial}{\partial t} + v_\parallel \frac{\partial}{\partial z} + \Omega \frac{\partial}{\partial \alpha} \right) \delta f  - \frac{e}{m} \left( \delta \bm E_\perp + \frac{\bm v \times \delta \bm B_\perp}{c} \right) \cdot \frac{\partial}{\partial \bm v} f_0 = 0 \; . \label{eq:vlasovlin}
\end{equation}
Noting the wave packet representation given by Eq. (\ref{eq:choruswp}) and using cylindrical coordinates in the velocity space $\bm v = (v_\perp, \alpha, v_\parallel)$ with $\alpha$ the gyrophase,
one can show
\begin{eqnarray}
& & \frac{e}{m} \left( \delta \bar{\bm E}_k + \frac{ \bm v \times  \delta \bar{\bm B}_k }{c} \right) \cdot \frac{\partial}{\partial \bm v} 
\nonumber \\ & & \hspace*{2em} = 
i \frac{e}{m} v_\perp \delta \bar E_k e^{i\alpha} \left[ \frac{k}{\omega}\frac{\partial}{\partial v_\parallel}
+ \left( 1 - \frac{k v_\parallel}{\omega} \right) \left( \frac{1}{v_\perp} \frac{\partial}{\partial v_\perp} + \frac{i}{v_\perp^2} \frac{\partial}{\partial \alpha} \right) \right]
\; , \label{eq:emforce}
\end{eqnarray}
where $\dot \alpha = \Omega$ (the positive definite electron cyclotron frequency) and we have denoted $\delta \bar E_k \equiv \left(\delta \bar{\bm E}_{\perp k}\right)_x$ for brevity. Thus, it is convenient to represent the 
supra-thermal electron response as
\begin{equation}
f (\bm v, z,t)  =  f_0 (\bm v, z) + \frac{1}{2} \sum_k \left(  e^{i S_k(z,t) + i \alpha} \delta \bar f_k (\bm v, z,t) + c.c. \right) \; , 
\label{eq:elef}
\end{equation}
with $f_0 (\bm v, z)$ representing the hot electron equilibrium distribution function. By direct substitution of Eqs. (\ref{eq:emforce}) and (\ref{eq:elef}) into 
Eq. (\ref{eq:vlasovlin}), it may be shown that
\begin{equation}
\delta \bar{f}_k =  \frac{e}{m} \frac{v_\perp \delta \bar E_k}{\left( k v_\parallel + \Omega - \omega \right)}
\left[ \frac{k}{\omega}\frac{\partial}{\partial v_\parallel}
+ \left( 1 - \frac{k v_\parallel}{\omega} \right) \frac{1}{v_\perp} \frac{\partial}{\partial v_\perp} \right]   f_0 \; . \label{eq:dbarfk}
\end{equation}
The presence of the resonant denominator in this expression shows that the Doppler shifted cyclotron
resonance condition is given by $\omega = \Omega + k v_\parallel$.
In order to calculate the wave-particle power transfer, we need to evaluate $\delta \bar{\bm J}_{h k} \cdot \delta \bar{\bm E}_{\perp k}^*$.
Noting that $\delta \bar{\bm J}_{h k \perp}  = - e \left \langle \bm v_\perp e^{i \alpha} \delta \bar f_k \right \rangle$, with angular brackets denoting velocity space
integration, and that
\begin{equation}
\bm v  \cdot \delta \bar{\bm E}_{\perp k}^* = - i v_\perp e^{-i\alpha} \delta \bar E_k^*\; , \label{eq:perpex}
\end{equation}
Eqs. (\ref{eq:dbarfk}) and (\ref{eq:perpex}) yield
\begin{eqnarray}
& & - \left.\frac{4 \pi i}{\omega} \frac{ \delta \bar{\bm J}_{h k} \cdot \delta \bar{\bm E}_{\perp k}^*}{|\delta \bar{\bm E}_{\perp k}(z,t)|^2}\right|_{k=K(z, \omega)}
= \left\langle \frac{4\pi e}{\omega} \frac{v_\perp }{2} \frac{\delta \bar E_k^* \delta \bar f_k}{|\delta \bar E_k|^2} \right\rangle
\nonumber \\
& & \hspace*{2em} = \frac{\omega_p^2}{n \omega} \left\langle \frac{v_\perp^2/2}{\Omega + 
k v_\parallel - \omega} 
\left[ \frac{k}{\omega} \frac{\partial f_0}{\partial v_{ \|}}+\left(1-\frac{k v_{ \|}}{\omega}\right) \frac{1}{v_{\perp}} \frac{\partial f_0}{\partial v_{\perp}} \right] \right\rangle \; , \label{eq:powerexchange}
\end{eqnarray}
where we have noted that $|\delta \bar{\bm E}_{\perp k}(z,t)|^2 = 2 |\delta \bar E_k|^2$.
In the next section, we will show that the unstable whistler wave packet growth and frequency modulation can be computed by 
means of the complex function
\begin{equation}
W(z, t, \omega)+i \Gamma(z, t, \omega) \equiv-\left.\frac{4 \pi i}{\omega \partial D_{w} / \partial \omega} \frac{ \delta \bar{\bm J}_{h k} \cdot \delta \bar{\bm E}_{\perp k}^*}{|\delta \bar{\bm E}_{\perp k}(z,t)|^2}\right|_{k=K(z, \omega)}  \; , \label{eq:WGamma}
\end{equation}
which, in the linear limit, is independent of time. 
Since the source region of supra-thermal electrons is localized near the equator,
the Earth's dipole magnetic field can be modeled as $B=B_e (1 + \xi z^2)$ \cite{Helliwell1967}, 
with $B_e$ the magnetic field strength at the equator and 
$\xi^{-1/2}$ the non-uniformity scale length. This model, thus, is equivalent to assuming $\Omega = \Omega_e (1 + \xi z^2)$
with $\Omega_e$ representing the (positive definite) electron cyclotron frequency at the equator. It is also commonly assumed
that the reference hot electron distribution function $f_0$ can be represented as a bi-Maxwellian \cite{Summers2012,Tao2014b}
\begin{equation}
f_0 = \frac{n_0}{(2\pi)^{3/2} w_{\parallel} w_{\perp}^2} \exp \left( - v_\perp^2/(2w_{\perp}^2) - v_\parallel^2/(2w_{\parallel}^2)  \right) \label{eq:f0eq}
\; ,
\end{equation}
where $n_0 = \zeta^2 n_e$, $w_{\parallel} = w_{\parallel e}$, $w_{\perp} = \zeta w_{\perp e}$, and 
$\zeta^{-2} = 1 + A \xi z^2/(1+ \xi z^2)$, with $A \equiv w_{\perp e}^2/w_{\parallel e}^2 - 1>0$ the anisotropy index computed at the equator,
where also all quantities with subscript ``$e$'' are evaluated. 

\begin{figure}[t]
	\begin{minipage}{0.5\linewidth}
	\begin{center}
		\resizebox{\textwidth}{!}{\includegraphics{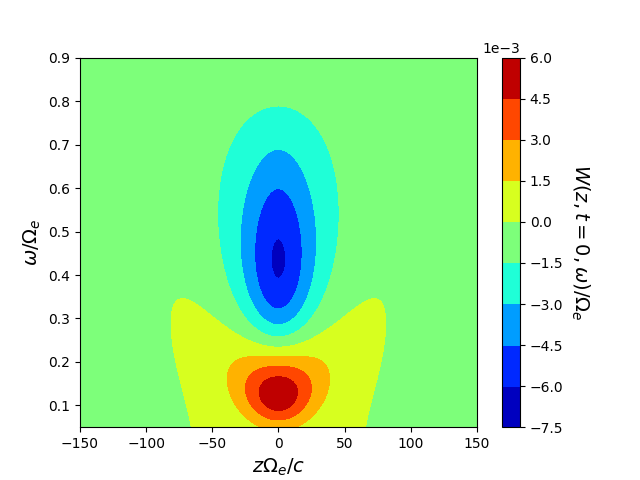}}
	\end{center} \end{minipage} \hfill \begin{minipage}{0.5\linewidth}
	\begin{center}
		\resizebox{\textwidth}{!}{\includegraphics{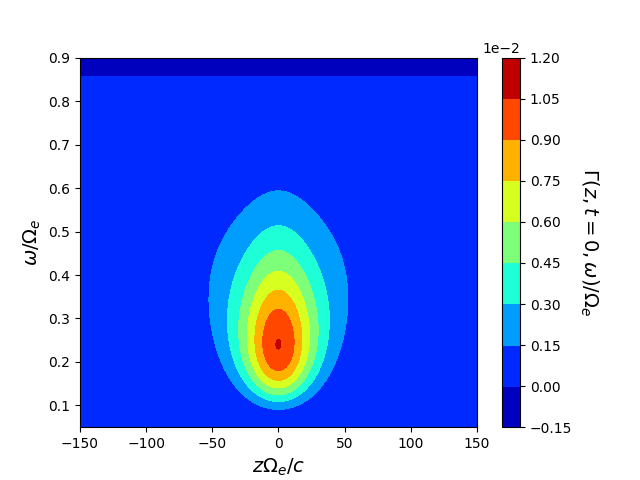}}
	\end{center} \end{minipage}
\vspace*{-2em}\newline\noindent(a) \hspace*{0.45\linewidth} (b)
\caption{Contour plots of $W(z, t=0, \omega)/\Omega_e$ (a) and $\Gamma(z, t=0, \omega)/\Omega_e$ (b) are shown
for normalized parameters $\omega_{p}/\Omega_e=5$, $n_e/n = 6 \times 10^{-3}$, $w_{\parallel e} = 0.2 c$, $w_{\perp e} = 0.53 c$,
$\xi = 8.62 \times 10^{-5} \Omega_e^2/c^2$.}
\label{fig:WGamma}
\end{figure}

By direct substitution of Eq. (\ref{eq:powerexchange}) into Eq. (\ref{eq:WGamma}), we obtain
\begin{eqnarray}
& & W(z, t, \omega)+i \Gamma(z, t, \omega)  =  \nonumber \\
& & \hspace*{2em} = \frac{\omega (\Omega - \omega)^2}{n \Omega} \left\langle \frac{v_\perp^2/2}{\Omega +  
k v_\parallel - \omega} 
\left[ \frac{k}{\omega} \frac{\partial f_0}{\partial v_{ \|}} - \frac{2}{v_\perp^2} \left(1-\frac{k v_{ \|}}{\omega}\right) f_0 \right] \right\rangle
 \; , \label{eq:WGamma4lin}
\end{eqnarray}
where, we have integrated by parts in $v_\perp$ and assumed $\omega^2/\omega_p^2 \ll 1$ in Eq. (\ref{eq:chorusdef}) to
explicitly write $\partial D_{w} / \partial \omega$ in compact form. Meanwhile, using Eq. (\ref{eq:f0eq}) 
and performing the velocity space integration, it is possible to write, with $t=0$ for the linear stage,
\begin{eqnarray}
& & W(z, 0, \omega)+i \Gamma(z, 0, \omega) = \frac{n_e(\Omega - \omega)^2}{n \Omega} \zeta^2 \left[ 1 - (A+1) \zeta^2   \right. \label{eq:WGammalin0} \\
& & \hspace*{2em} + (A+1) \zeta^2 \frac{(\Omega - \omega)}{\sqrt{2}|k| w_{\parallel e}} Z\left( \frac{\omega-\Omega}{\sqrt{2}|k| w_{\parallel e}} \right) \left.- \frac{\Omega}{\sqrt{2}|k| w_{\parallel e}} 
Z\left( \frac{\omega-\Omega}{\sqrt{2}|k| w_{\parallel e}} \right) \right] \; , \nonumber
\end{eqnarray}
where $Z(x) = \pi^{-1/2} \int_{-\infty}^\infty e^{-y^2}/(y-x) dy$ is the plasma
dispersion function. Recalling $A+1 - \zeta^{-2} = A/(1+\xi z^2)$, Eq. (\ref{eq:WGammalin0}) can be rewritten as
\begin{eqnarray}
& & W(z, 0, \omega)+i \Gamma(z, 0, \omega) = \frac{n_e(\Omega - \omega)^2}{n \Omega} \zeta^4 \left[ - \frac{A}{1+\xi z^2} \right. \nonumber \\
& &  \hspace*{4em} \left. +  \frac{A\Omega_e - (A+1)\omega}{\sqrt{2}|k| w_{\parallel e}} Z\left( \frac{\omega-\Omega}{\sqrt{2}|k| w_{\parallel e}} \right) \right] \; . \label{eq:WGammalin}
\end{eqnarray}
Thus, $\omega/\Omega_e = A/(A+1)$ is the frequency where wave particle power exchange with hot electrons
changes sign and  becomes a damping \cite{Kennel1966b}. Meanwhile, Eq. (\ref{eq:WGammalin}) also shows
that $W(z, 0, \omega)$ and $\Gamma(z, 0, \omega)$ scale as $\zeta^4$. Thus, the 
wave-particle interaction with supra-thermal electrons is characterized by the length scale $\sim (A\xi)^{-1/2}$, which
competes with and actually takes over the non-uniformity due to the ambient magnetic field $B = B_e (1 + \xi z^2)$ \cite{Helliwell1967}
already at moderate values of $A$. For this reason and for the sake of simplicity, it may be convenient,
especially in theoretical/analytical studies \cite{Zonca2021}, to assume  
a non-uniform source of hot electrons as in Eq. (\ref{eq:f0eq}) with $\zeta^{-2} \simeq 1 + A \xi z^2$, localized about
the equator, neglecting, meanwhile, magnetic field non-uniformity. 
Contour plots of the (linear) functions $W(z, t=0, \omega)$ and $\Gamma(z, t=0, \omega)$ are given in
Fig. \ref{fig:WGamma}, for normalized parameters $\omega_{p}/\Omega_e=5$, $n_e/n = 6 \times 10^{-3}$, $w_{\parallel e} = 0.2 c$, $w_{\perp e} = 0.53 c$,
$\xi = 8.62 \times 10^{-5} \Omega_e^2/c^2$ \cite{Tao2014b}.

\subsection{Whistler wave packet propagation}
\label{sec:whistlerwp}

As noted in Sec. \ref{sec:whistler}, ``due to the large separation between wavelength and magnetic field line scale length, we can describe whistler waves as wave packets propagating in a weakly nonuniform medium'' \cite{Bernstein75,Kravtsov90,Zonca2021}. In particular, by letting 
\begin{equation}
\delta \bar{\bm E}_{\perp k}(z,t) = \hat{\bm e} |\delta \bar{\bm E}_{\perp k}(z,t)| e^{i \varphi_k (z,t)} \; , \label{eq:ampphase}
\end{equation}
with $\hat{\bm e}$ the polarization vector defined such $\hat{\bm e} \cdot \hat {\bm z} = 0$ and $\hat{\bm e} \cdot \hat{\bm e}^* = 1$;
and introducing 
\begin{equation}
I_k(z,t) \equiv \left|\partial D_w/\partial k\right| |\delta \bar{\bm E}_{\perp k}(z,t)|^2 \; , \label{eq:waveaction}
\end{equation}
the evolution equation for $I_k(z,t)$ is 
\begin{equation}
\left( \frac{\partial}{\partial t} + v_{gk} \frac{\partial}{\partial z} \right) I_k(z,t) = 2 \gamma_k I_k(z,t)  = 2 \Gamma (z,t,\omega) I_k(z,t) \; , \label{eq:actionevolve}
\end{equation}
where $v_{gk} = - (\partial D_w/\partial \omega_k)^{-1}\partial D_w/\partial k$ is the wave-packet group velocity and
\begin{equation}
\gamma_k = - \frac{D_{Ak}^1}{\partial D_w/\partial \omega_k}
\end{equation}
represents the wave packet driving rate due to the supra-thermal electrons. 
In fact, noting $\partial_z\partial_k D_w = 0$, which allows us to identify the adjoint of $D_w$ with its complex conjugate \cite{McDonald1988},
\begin{equation}
D_{A k}^1 = \mathbb I{\rm m} \left( \frac{4\pi i}{\omega_k} \frac{ \delta \bar{\bm J}_{h k} \cdot \delta \bar{\bm E}_{\perp k}^*}{|\delta \bar{\bm E}_{\perp k}(z,t)|^2} \right)\; 
\label{eq:antiHD}
\end{equation}
is the anti-Hermitian part of the (perturbed) dielectric constant. Equation (\ref{eq:actionevolve}) is derived by taking the scalar product of Eq. (\ref{eq:master}) with 
$\delta \bar{\bm E}^*_{\perp k}(z,t)$ and then treating it as wave equation in a slowly-varying weakly non-uniform medium \cite{Bernstein75,Kravtsov90}.
Meanwhile, the phase shift $\varphi_k (z,t)$ is given by
\begin{equation}
 \left( \frac{\partial}{\partial t} + v_{gk} \frac{\partial}{\partial z} \right) \varphi_k (z,t) = \frac{D_{R k}^1}{\partial D_w/\partial \omega_k}  = - W (z,t,\omega)\; ,  \label{eq:phaseevolve}
\end{equation}
where 
\begin{equation}
D_{R k}^1 = \mathbb R{\rm e} \left( \frac{4\pi i}{\omega_k} \frac{ \delta \bar{\bm J}_{h k} \cdot \delta \bar{\bm E}_{\perp k}^*}{|\delta \bar{\bm E}_{\perp k}(z,t)|^2} \right) \; , 
\label{eq:realHD}
\end{equation}
is the Hermitian part of the (perturbed) dielectric constant. Thus, all relevant nonlinear physics is included in the wave-particle interactions between
whistler waves and supra-thermal electrons. More precisely, it is all described by the 
two functions $W(z, t, \omega)$ and $\Gamma(z, t, \omega)$ introduced in Eq. (\ref{eq:WGamma}), derived in Sec. \ref{sec:whistlerinst} and then specialized
to Eqs. (\ref{eq:WGamma4lin}) and (\ref{eq:WGammalin}) in the linear limit. We will see in Sec. \ref{sec:phasespace} that Eq. (\ref{eq:WGamma4lin}) still holds in the nonlinear
case. 

In the light of the discussion above, Eqs. (\ref{eq:actionevolve}) and (\ref{eq:phaseevolve}) describe both linear and nonlinear evolution of the whistler wave spectrum excited by supra-thermal electrons once $W(z, t, \omega)$ and $\Gamma(z, t, \omega)$ are given. These equations are solved by the method of characteristics, 
where, given the lowest order WKB dispersion relation, Eq. (\ref{eq:chorusdisp}), all fields with a subscript $k = K(z,\omega)$ must be interpreted as functions in the 
$(z,\omega)$ space. That is,  $I_k(z,t)|_{k=K(z,\omega)} \equiv I (z, t, \omega)$, $\varphi_k(z,t)|_{k=K(z,\omega)} \equiv \varphi (z, t, \omega)$ and the group velocity
$v_{g k}(z)|_{k=K(z,\omega)} = v_{g}(z,\omega)$. 
Defining a wave packet time variable
\begin{equation}
T_\omega(z) \equiv \int_0^z  \frac{dz'}{v_{g}(z',\omega)} \; , \label{eq:Tktrans}
\end{equation}
and its inverse 
\begin{equation}
T^{-1}_\omega T_\omega(z) \equiv z  \; ; \label{eq:Tktransinv}
\end{equation}
the initial position, $z_0$, at $t=0$ of a wave packet located at position $z$ at time $t$ is
then given by $T_\omega(z) - T_\omega(z_0) = t$; or, equivalently
\begin{equation}
z_0 (z,t) = T^{-1}_\omega T_\omega(z_0)  = T_\omega^{-1}(T_\omega(z)- t)  \; . \label{eq:z0}
\end{equation}
It is then straightforward to show that
\begin{eqnarray}
\partial_t \left[T_\omega^{-1}(T_\omega(z)- t)\right]  & = & - v_g \left( T_\omega^{-1}(T_\omega(z)- t) , \omega \right) \; , \nonumber \\
\partial_z \left[T_\omega^{-1}(T_\omega(z)- t)\right]  & = & v_g \left( T_\omega^{-1}(T_\omega(z)- t) , \omega \right)/v_g(z,\omega) \; . \label{eq:Twdery}
\end{eqnarray}
Given Eq. (\ref{eq:Twdery}), one may verify by direct substitution that
\begin{eqnarray}
& & I (z, t, \omega) = I_{\omega 0} \left( T_\omega^{-1}(T_\omega(z)- t) \right) \nonumber \\
& & \hspace*{2em} \times \exp \left( 2 \int_{T_\omega^{-1}(T_\omega(z)- t)}^z \frac{dz'}{v_{g}(z',\omega)} \Gamma(z',t-T_\omega(z)+T_\omega(z'),\omega) \right) 
\; , \label{eq:Ikevolve}
\end{eqnarray}
and
\begin{eqnarray}
& & \varphi (z, t, \omega)  =  \varphi_{\omega 0} \left(T_\omega^{-1}(T_\omega(z)- t)\right) \nonumber \\
& & \hspace*{2em}  - \int_{T_\omega^{-1}(T_\omega(z)- t)}^z \frac{dz'}{v_{g}(z',\omega)} W(z',t-T_\omega(z)+T_\omega(z'),\omega) \; , \label{eq:phikevolve}
\end{eqnarray}
are solutions of Eqs. (\ref{eq:actionevolve}) and (\ref{eq:phaseevolve}).
Here, $I_{\omega 0}(z) = I(z,0,\omega)$ and  $\varphi_{\omega 0}(z) = \varphi(z,0,\omega)$ are the initial conditions for the considered wave packet. 
In fact, using Eq. (\ref{eq:Twdery}), one can show that, assuming $I_{\omega 0} (z) =$ const for simplicity,
\begin{eqnarray}
\frac{\partial_t I}{2 I} & = & \Gamma \left( T_\omega^{-1}(T_\omega(z)- t) , 0 , \omega \right) 
\nonumber \\ & & + \int_{T_\omega^{-1}(T_\omega(z)- t)}^z \frac{dz'}{v_{g}(z',\omega)} \partial_t \Gamma (z',t-T_\omega(z)+T_\omega(z'),\omega)  \; ;  \label{eq:deryid1}
\end{eqnarray}
\begin{eqnarray}
\frac{\partial_z I}{2 I} & = &  \frac{1}{v_g(z,\omega)} \left[ \Gamma \left( z , t , \omega \right) - \Gamma \left( T_\omega^{-1}(T_\omega(z)- t) , 0 , \omega \right) \right] \nonumber \\ & & + \int_{T_\omega^{-1}(T_\omega(z)- t)}^z \frac{dz'}{v_{g}(z',\omega)} \partial_z \Gamma (z',t-T_\omega(z)+T_\omega(z'),\omega) \; . \label{eq:deryid2}
\end{eqnarray}
Analogous equations can be written for $\partial_t \varphi$ and $\partial_z \varphi$. 
Again, Eqs. (\ref{eq:Ikevolve}) and (\ref{eq:phikevolve}) completely describe the linear and nonlinear evolution of the system, 
once $W(z, t, \omega)$ and $\Gamma(z, t, \omega)$ are given. 
Noting that $\partial_t \Gamma \left( z , t , \omega \right) = 0$ in the linear limit, Eq. (\ref{eq:deryid1}), in particular, has two major implications: (i) the
whistler instability is of convective type (it does not grow exponentially at the same rate everywhere in the time asymptotic limit); (ii) chorus chirping
is a nonlinear process. In fact, assuming that chirping occurs, the behavior of $I(z,t,\omega)$ at a given frequency and position should first
increase in time and then decrease, reaching a maximum during the time interval corresponding to the chorus element 
crossing the considered frequency at the given position.
In order for this to be possible in the linear limit; \idest, for $\partial_t \Gamma \left( z , t , \omega \right) = 0$, 
we should have $\Gamma \left( T_\omega^{-1}(T_\omega(z)- t) , 0, \omega \right) = 0$ at the considered frequency and given position, 
which clearly cannot be the case. 

\begin{figure}[t]
	\begin{minipage}{0.5\linewidth}
	\begin{center}
		\resizebox{\textwidth}{!}{\includegraphics{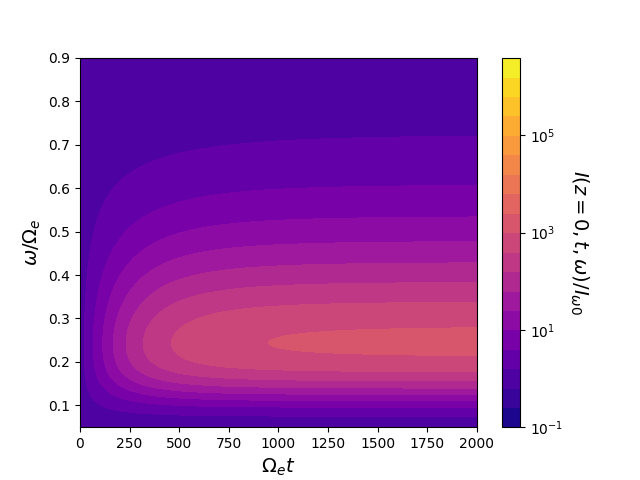}}
	\end{center} \end{minipage} \hfill \begin{minipage}{0.5\linewidth}
	\begin{center}
		\resizebox{\textwidth}{!}{\includegraphics{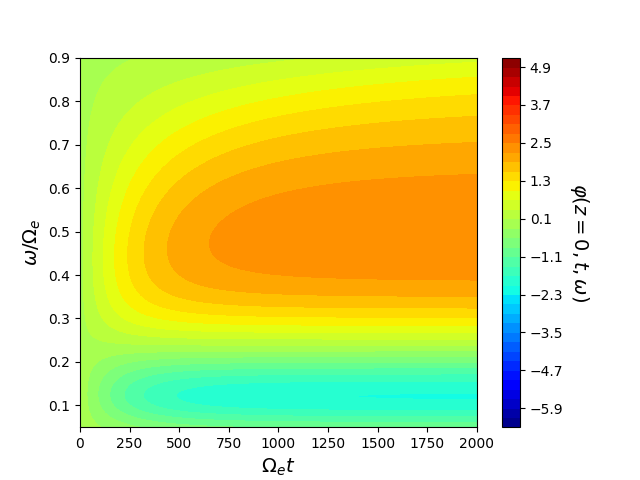}}
	\end{center} \end{minipage}
\vspace*{-2em}\newline\noindent(a) \hspace*{0.45\linewidth} (b) \newline 
	\begin{minipage}{0.5\linewidth}
	\begin{center}
		\vspace*{.2em}\resizebox{\textwidth}{!}{\includegraphics{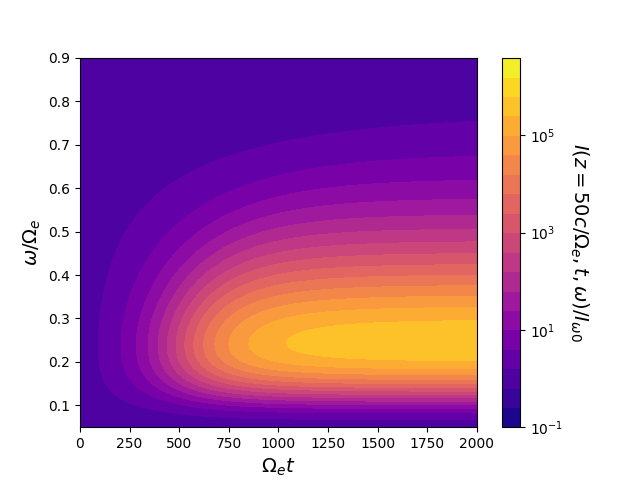}}
	\end{center} \end{minipage} \hfill \begin{minipage}{0.5\linewidth}
	\begin{center}
		\vspace*{.2em}\resizebox{\textwidth}{!}{\includegraphics{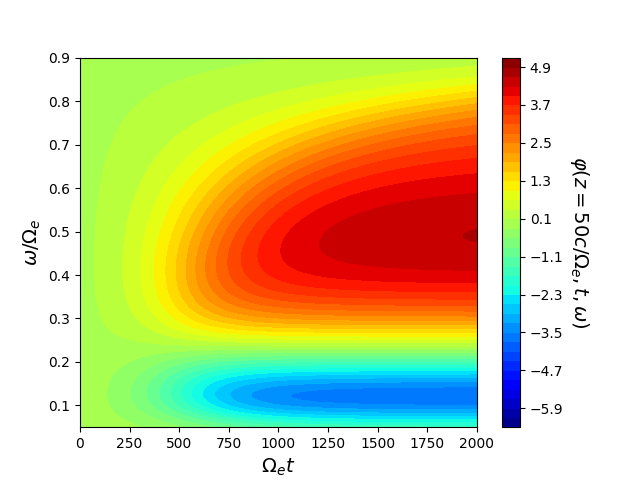}}
	\end{center} \end{minipage}
\vspace*{-2em}\newline\noindent(c) \hspace*{0.45\linewidth} (d) \vspace*{0em}
\caption{Contour plots in the $(\omega,t)$ plane that illustrate the linearized time evolution of $I(z,t,\omega)/I_{\omega 0}$ 
and $\varphi(z,t,\omega)$ assuming $I_{\omega 0} = {\rm const.}$ and $\varphi_{\omega 0} = 0$.  Physical parameters
are those of Fig. \ref{fig:WGamma}. Top panels (a) and (b) refer to $z=0$; while panels (c) and (d) are their analogue 
at $z=50 c/\omega_e$.}
\label{fig:Ivarphi_z0}
\end{figure}

\begin{figure}[t]
	\begin{minipage}{0.5\linewidth}
	\begin{center}
		\resizebox{\textwidth}{!}{\includegraphics{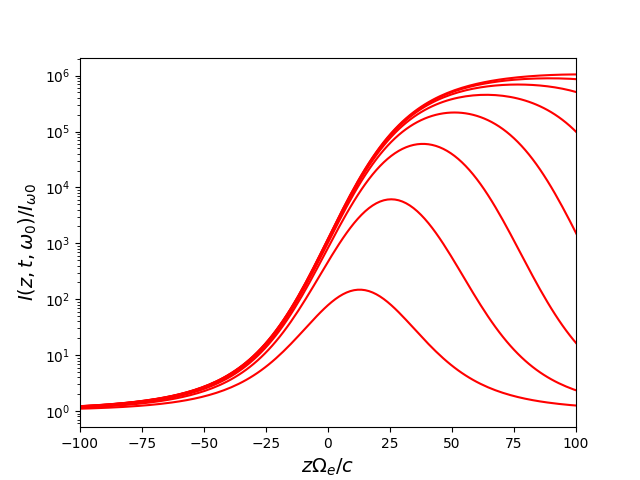}}
	\end{center} \end{minipage} \hfill \begin{minipage}{0.5\linewidth}
	\begin{center}
		\resizebox{\textwidth}{!}{\includegraphics{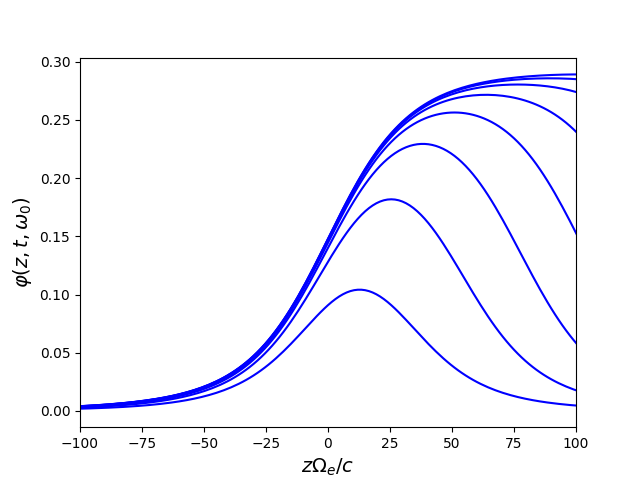}}
	\end{center} \end{minipage}
\vspace*{-2em}\newline\noindent(a) \hspace*{0.45\linewidth} (b) 
\caption{Snapshots at times $\Omega_e t = 250, 500, ..., 2000$ that illustrate the linearized time evolution of $I(z,t,\omega_0)/I_{\omega 0}$ 
(a) and $\varphi(z,t,\omega_0)$ (b) assuming $I_{\omega 0} = {\rm const.}$ and $\varphi_{\omega 0} = 0$.  Increasingly higher-valued curves
correspond to later times. Physical parameters
are those of Fig. \ref{fig:WGamma}.}
\label{fig:Ivarphi_omega0}
\end{figure}

In order to illustrate the linear spatiotemporal structures of the whistler wave packets excited by supra-thermal electrons,
Fig. \ref{fig:Ivarphi_z0} shows contour plots in the $(\omega,t)$ plane of $I(z,t,\omega)/I_{\omega 0}$ and 
$\varphi(z,t,\omega)$ assuming $I_{\omega 0} = {\rm const.}$ and $\varphi_{\omega 0} = 0$.  For simplicity,
only wave packets with positive $k$ and correspondingly positive group velocity are considered, with $W(z,0,\omega)$
and $\Gamma(z,0,\omega)$ given by Eq. (\ref{eq:WGammalin}).
Top panels (a) and (b) in Fig. \ref{fig:Ivarphi_z0} refer to $z=0$; \idest\, to the fluctuation intensity and phase at the
equator, where the supra-thermal electron source is strongest. Meanwhile, panels (c) and (d) are, respectively, the analogue 
of panels (a) and (b) at $z=50 c/\omega_e$, where the hot electron source has significantly decayed but the wave packet
has traveled though the whole source region. 
Finally, for $\omega = \omega_0 \simeq 0.241 \Omega_e$, corresponding to the strongest growing frequency
at the equator in Fig. \ref{fig:WGamma}, Fig. \ref{fig:Ivarphi_omega0} shows snapshots of the spatial
structure of both $I(z,t,\omega_0)/I_{\omega 0}$ 
(a) and $\varphi(z,t,\omega_0)$ (b) assuming $I_{\omega 0} = {\rm const.}$ and $\varphi_{\omega 0} = 0$.
After the initial peaking of both $I(z,t,\omega_0)$ and $\varphi(z,t,\omega_0)$ at $z\simeq 0$, where the growth rate is strongest, the
propagation of the whistler wave packet at the group velocity along the magnetic field line becomes increasingly more evident\footnote{Recall that, here,
for the sake of simplicity, we are considering only positive $k$, corresponding to wave packets propagating from negative to positive $z$.}.
For sufficiently large $z$, such that it is outside the ``hot electron'' source region (cf. Sec. \ref{sec:chorusobs}), Eqs. (\ref{eq:Ikevolve}) and (\ref{eq:phikevolve}) 
suggest that spatiotemporal dependences can be only on $T_\omega^{-1}(T_\omega(z)- t)$. Time asymptotically, when
wave packets have traveled all the way from $z \rightarrow - \infty$ to $z \rightarrow \infty$, $I(z,t,\omega_0)$ and $\varphi(z,t,\omega_0)$
reach a constant shape, which depends on $\omega_0$ only.

\section{Phase space transport and nonlinear dynamics}
\label{sec:phasespace}

In this section, we introduce the nonlinear dynamics starting from the basic kinematic description of 
single particle motion and wave particle trapping in Sec. \ref{sec:singlep}. 
A phase space hole is created during this process near resonance: that is, a phase space density difference 
between the center of the resonant structure and the ``unperturbed surrounding''. 
The phase space structure is called ``phase space hole/clump'' depending on the sign of Eq. (\ref{eq:Deltaf0eq}).
The dynamic description 
of phase space nonlinear behaviors is given in Sec. \ref{sec:smallamp}, 
where we focus on the nonlinear particle response without fast temporal or spatial dependences, 
which correspond to the self-interaction of the wavenumber of interest with itself. 
This nonlinear response obeys to the evolution equation, Eq. (\ref{eq:renorm}), which
may look like a ``quasilinear''  description at first glance but is, indeed, much more general
as will be shown and discussed in the following. For the more theoretically inclined readers,
we will delineate the close connection of the present approach with 
the Dyson equation and the so-called ``principal series'' of secular terms.
Conservation properties of the adopted reduced description are discussed in
Sec. \ref{sec:conserve}, where we also analyze the different nonlinear dynamic behaviors
that are possibly described within the present theoretical framework, anticipating
some of the qualitative features of chorus excitation, which are addressed in Sec. \ref{sec:chorus}.

\subsection{Single particle motion and wave particle trapping}
\label{sec:singlep}

\begin{figure}[t]
	\begin{center}
		\resizebox{\textwidth}{!}{\includegraphics{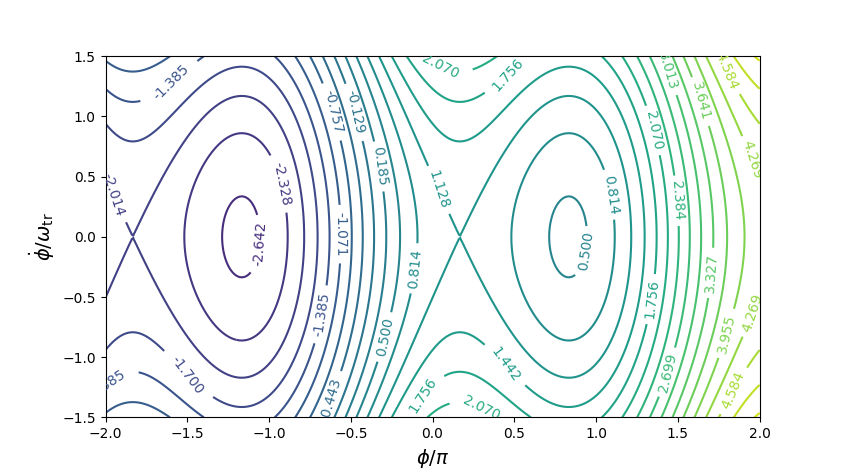}}
	\end{center} 
\caption{Contour plot of Eq. (\ref{eq:ddphi2}) in the $(\phi,\dot\phi)$ phase space showing structure formation
due to wave particle trapping for $R=1/2$.}
\label{fig:wavetrap}
\end{figure}

The equations of motion of single particles (electrons) interacting with the whistler wave packet 
described in Sec. \ref{sec:choruslin} are the natural starting point for analyzing nonlinear dynamics.
Noting Eq. (\ref{eq:emforce}) to extract the components of the Lorentz force, it is possible to write
\begin{eqnarray}
& & \dot{v}_{\perp}= \frac{v_{\|} v_{\perp}}{2 \Omega} \frac{d \Omega}{d z} - \frac{e}{m}\left(\frac{k v_{\|}}{\omega}-1\right) \delta \bar E_k \sin \phi
\; , \nonumber \\ 
& & \dot{v}_{\|}= - \frac{v_{\perp}^2}{2 \Omega} \frac{d \Omega}{d z} + \frac{e}{m} \frac{k v_{\perp}}{\omega} \delta \bar E_k  \sin \phi \; , \nonumber \\ 
& & \dot{\phi}=\Omega-\omega+k v_{\|} - \frac{e}{m v_{\perp}}\left(\frac{k v_{\|}}{\omega}-1\right) \delta \bar E_k  \cos \phi
\; , \label{eq:singlep}
\end{eqnarray}
where $\phi \equiv \alpha + S_k$, and we recalled that $\dot \alpha = \Omega$ (the positive definite electron cyclotron frequency) 
and $\delta \bar E_k \equiv \left(\delta \bar{\bm E}_{\perp k}\right)_x$ for brevity, which is assumed as real without loss of
generality. Noting that 
the considered wave packet is right handed circularly polarized, and $\left(\delta \bar{\bm B}_{\perp k}\right)_y 
= (kc/\omega) \left(\delta \bar{\bm E}_{\perp k}\right)_x$, Eqs. (\ref{eq:singlep}) are readily cast in the standard form
adopted by \cite{Dysthe1971,Nunn1974,Inan1978,Vomvoridis1979,Omura2008,Shklyar2009,Albert2012} and, more recently,
in the brief review by \cite{Tao2020}. In terms of the magnetic moment, $\mu = v_\perp^2/2$, and the energy per
unit mass, ${\cal E} = v^2/2$, Eqs. (\ref{eq:singlep}) can be rewritten as
\begin{eqnarray}
& & \dot{\mu} = - \frac{e}{m}\left(\frac{k v_{\|}}{\omega}-1\right) \left( \frac{2 \mu}{B} \right)^{1/2} \delta \bar E_k \sin \phi
\; , \nonumber \\ 
& & \dot{\cal E} = \frac{e}{m}  \left( 2 \mu B \right)^{1/2} \delta \bar E_k  \sin \phi \; , \nonumber \\ 
& & \dot{\phi}=\Omega-\omega+k v_{\|} 
\; , \label{eq:singlepen}
\end{eqnarray}
where, in the latter equation, we have dropped the negligible contribution to $\dot \phi$ due $\propto \delta \bar E_k$ 
(cf., \exgra, any of the above references in this subsection). The resonance velocity is given by $\dot \phi = 0$ and, 
thus, 
\begin{equation}
v_{r} = \frac{\omega - \Omega}{k} \; . \label{eq:vr}
\end{equation}
Equations (\ref{eq:vg}) and (\ref{eq:vr}) show that the resonant velocity of the considered whistler wave packet 
has opposite sign with respect to the corresponding phase and group velocities. Noting that $\dot \phi = k ( v_\parallel - v_r)$,
it is readily shown that, near resonance,
\begin{equation}
\ddot \phi = k \frac{d}{dt} ( v_\parallel - v_r) \; , \label{eq:ddphi0}
\end{equation}
where the total time derivative along the moving wave packet is expressed as
\begin{equation}
\frac{d}{dt} = \frac{\partial}{\partial t} + v_g \frac{\partial}{\partial z} \; . \label{eq:ddt}
\end{equation}
Using Eqs. (\ref{eq:vg}) and (\ref{eq:vr}) along with the equations of motion, Eqs. (\ref{eq:singlep}),
and assuming constant $\omega_{p}^2$ along the magnetic field line,
one can show, after some lengthy but straightforward algebra (cf., \exgra, \cite{Vomvoridis1982,Omura2008}),
\begin{equation}
\ddot \phi = \omega_{\rm tr}^2 \left( \sin \phi - R \right) \; , \label{eq:ddphi1}
\end{equation}
where \cite{Vomvoridis1982,Omura2008}
\begin{equation}
R=\frac{1}{\omega_{\rm tr}^{2}}\left[\left(1-\frac{v_{r}}{v_{g}}\right)^{2} \frac{\partial \omega}{\partial t}+\left(\frac{k v_{\perp}^{2}}{2 \Omega}-\frac{3}{2} v_{r}\right) \frac{\partial \Omega}{\partial z}\right] \; , \label{eq:R}
\end{equation}
and $\omega_{\rm tr}^2 = k v_\perp (e/m) k  \delta \bar E_k/\omega = k v_\perp (e \delta \bar B_{k y}/ mc)$\footnote{Note that, with the
present notation and definitions, $\omega_{\rm tr}^2$ is positive definite. Should this not be the case, it would be sufficient to redefine
$\phi$ introducing a phase shift of $\pi$ to recover the physical meaning of Eq. (\ref{eq:ddphi1}).}. The derivation of $R$ expression 
is based on kinematics and wave dispersive properties only; and follows from direct calculation of 
$k d v_r/dt$ \cite{Vomvoridis1982,Omura2008}. For $R=0$, Eq. (\ref{eq:ddphi1}) is a nonlinear pendulum equation
with $\omega_{\rm tr}$ giving the frequency of small amplitude oscillations. For $|R| < 1$,  structures are formed
in the $(\phi,\dot\phi)$ phase space due to wave particle trapping. This is shown in Fig. \ref{fig:wavetrap}
for $R=1/2$. Meanwhile, no structure formation is possible for $|R| >1$, as can be seen 
integrating once Eq. (\ref{eq:ddphi1}), yielding
\begin{equation}
(1/2) \dot \phi^2 + \omega_{\rm tr}^2 \left( \cos \phi + R \phi \right) = {\rm const} \; . \label{eq:ddphi2}
\end{equation}
We will come back to the importance of this condition in Sec. \ref{sec:chorus}, since $R$ is evidently
connected with the frequency chirping of the whistler wave packet from its definition, Eq. (\ref{eq:R}).
Here, we further note that, unlike the usual nonlinear pendulum case, the separatrix 
identifying phase space oscillations for $|R|<1$ has one single X-point. Again, we will discuss the
important implications of this fact in Sec. \ref{sec:chorus}.

Near resonance, Eqs. (\ref{eq:singlepen}) imply that
\begin{equation}
\dot \mu B \omega = \Omega \dot{\cal E} \; , \;\;\;\;\; \Rightarrow  \;\;\;\;\; \omega d\mu = (\Omega/B) d {\cal E} \; . 
\label{eq:dmudE}
\end{equation}
Thus, the increase/decrease of perpendicular energy due to wave-particle interaction is larger than the
corresponding increase/decrease of total energy for $0 < \omega < \Omega$. Said differently, an increase in magnetic moment and
energy of a resonant particle is necessarily accompanied by a decrease in the parallel energy and {\em vice versa}.
Considering, \exgra, a rising tone chorus element (cf. Sec. \ref{sec:chorus} for details) and, for simplicity,
only wave packets with positive $k$ as we did in Sec. \ref{sec:choruslin}, resonant velocity, Eq. (\ref{eq:vr}), is negative
and becomes progressively less negative (resonant velocity increases). Thus, a {\em phase locked}
resonant particle, such as those that are {\em deeply trapped} near O-points of structures in the 
$(\phi,\dot\phi)$ phase space in Fig. \ref{fig:wavetrap}, looses parallel energy in the chirping process and,
at the same time, gains total energy as well as magnetic moment. It can be shown that a {\em phase space hole}
is created during this process near resonance. In fact, by rewriting Eq. (\ref{eq:f0eq}) as
\begin{equation}
f_{0}=\frac{n_{e}}{(2 \pi)^{3 / 2} w_{\| e} w_{\perp e}^{2}} \exp \left(-\mathcal{E} / w_{\| e}^{2}+A \mu B_{e} / w_{\perp e}^{2}\right) \label{eq:f0eqemu}
\; ,
\end{equation}
we can evaluate the phase space density difference, $\Delta f_0$, between the center of the resonant structure and the
``unperturbed surrounding''. The center of the resonant structure is characterized by $f_{0}$ at the
original $\mathcal{E}$ and $\mu$. Meanwhile the unperturbed surrounding is characterized by $f_{0}$ at the
$\mathcal{E}+\Delta \mathcal{E}$ and $\mu + \Delta \mu$. Thus, using Eq. (\ref{eq:dmudE}) 
and recalling $A = w_{\perp e}^{2}/w_{\parallel e}^{2} - 1$ \cite{Nunn1974,Vomvoridis1982,Omura2008,Tao2020}
\begin{eqnarray}
\Delta f_{0} & = &f_{0} ({\cal E}, \mu) - f_{0} ({\cal E} + \Delta {\cal E}, \mu + \Delta \mu) \; , \nonumber \\
& \propto &  \frac{\Delta \mathcal{E}}{w_{\| e}^{2}} - A B_{e} \frac{\Delta \mu}{w_{\perp e}^{2}} =  - A B_{e} \frac{\Delta \mu}{w_{\parallel e}^{2}} 
\left( \frac{A}{A+1} - \frac{\omega}{\Omega_e} \right)
\label{eq:Deltaf0eq} \; .
\end{eqnarray}
Depending on the negative/positive sign of Eq. (\ref{eq:Deltaf0eq}), the resonant phase space structure is called ``phase space hole/clump''.
This demonstrates that phase space hole is formed near resonance for a rising tone chorus element
characterized by $\Delta \mu > 0$ [cf. discussion between Eqs. (\ref{eq:dmudE}) and (\ref{eq:f0eqemu})], which tends to disappear when
the whistler wave instability is lost for $\omega/\Omega_e = A/(A+1)$, consistent with Eq. (\ref{eq:WGammalin}) and Fig. \ref{fig:WGamma}(b).
The same argument can be used to show that a {\em phase space clump}
occurs near resonance for a falling tone chorus element.

\subsection{Small amplitude expansion and renormalized electron response}
\label{sec:smallamp}

\begin{figure}[t]
	\begin{minipage}{0.5\linewidth}
	\begin{center}
		\resizebox{0.6\textwidth}{!}{\includegraphics{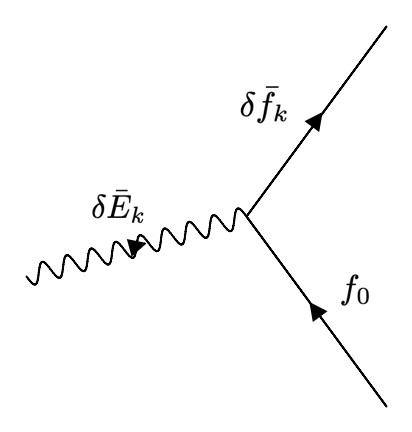}}
	\end{center} \end{minipage} \hfill \begin{minipage}{0.5\linewidth}
	\begin{center}
		\resizebox{\textwidth}{!}{\includegraphics{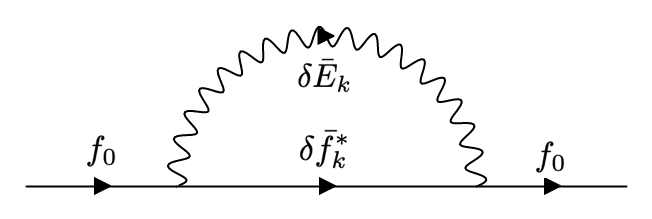}}
	\end{center} \end{minipage}
\vspace*{-2em}\newline\noindent(a) \hspace*{0.45\linewidth} (b) 
\caption{Schematic illustration of Eq. (\ref{eq:dbarfk})
borrowing Feynman diagrams with suitably adapted rules (a)
 \cite{Zonca2015, Zonca2015b, Chen2016, Zonca2021};
and modification of the original equilibrium defined in Eq. (\ref{eq:renorm0}) represented as simple {\em loop diagram} (b).}
\label{fig:feynman}
\end{figure}

\begin{figure}[t]
	\begin{center}
		\resizebox{\textwidth}{!}{\includegraphics{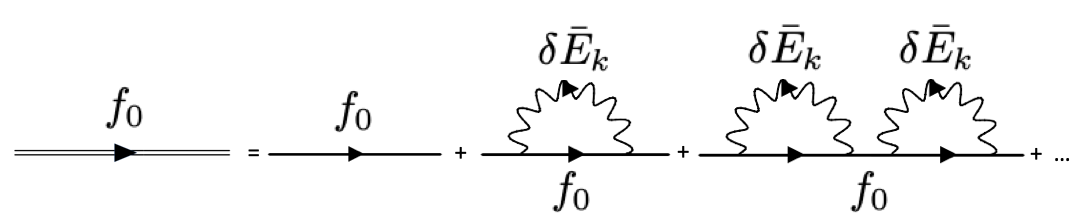}}
	\end{center}
\caption{Schematic illustration of the {\em Dyson series} iterative solution
of Eqs. (\ref{eq:dbarfk}) and (\ref{eq:renorm}).}
\label{fig:dyson}
\end{figure}

Addressing nonlinear dynamics and phase space transport underlying the evolution
of whistler wave packet excited by supra-thermal electrons requires analyzing the 
self-consistent energetic electron response in the presence of a small but finite fluctuation level.
In Sec. \ref{sec:choruslin}, we have noted that the effect of supra-thermal electrons can
be treated as a perturbation, with wave dispersive properties remaining those of 
a right-hand circular polarized whistler wave propagating parallel to the Earth's
magnetic field. The convective amplification of the whistler wave packet by the energetic
electron source, localized nearby the equator, suggests that the fluctuation level remains
small (\exgra, $|\delta \bar{\bm B}_{\perp k}/B| \ll 1$) during the nonlinear evolution and, thus,
that the self-consistent electron response may be computed assuming a small amplitude expansion.
That is, we assume that the electron distribution function can be formally written as an asymptotic 
series
\begin{equation}
f = f^{(0)} + f^{(1)} + f^{(2)} + \ldots \; , \label{eq:fasy}
\end{equation}
where the superscripts $(n)$ denote the power at which fluctuation amplitude enters in each term.
The first two terms are those that we analyzed in Sec. \ref{sec:choruslin}. More precisely, noting
Eq. (\ref{eq:elef}), $f^{(0)} = f_0 (\bm v, z)$ is the reference equilibrium distribution function, while
$f^{(1)}$ corresponds to terms $\propto \delta \bar f_k (\bm v, z,t)$ and c.c. Noting the form of
the nonlinear Vlasov equation
\begin{equation}
\left( \frac{\partial}{\partial t} + v_\parallel \frac{\partial}{\partial z} + \Omega \frac{\partial}{\partial \alpha} \right) f  - \frac{e}{m} \left( \delta \bm E_\perp + \frac{\bm v \times \delta \bm B_\perp}{c} \right) \cdot \frac{\partial}{\partial \bm v} f = 0 \; ; \label{eq:vlasov}
\end{equation}
and the structure of the nonlinear coupling term that can still be represented by Eq. (\ref{eq:emforce}), it can be
shown that the natural extension of equation Eq. (\ref{eq:elef}) is
\begin{eqnarray}
f^{(2)}  & = & \frac{1}{2}  \sum_{k,k'} \left(  e^{i S_k(z,t) + i S_{k'}(z,t) + 2 i \alpha} \delta \bar f_{k+k'} (\bm v, z,t) + c.c. \right) \nonumber \\
& & +  \frac{1}{2}  \sum_{k,k'} \left(  e^{i S_k(z,t) - i S_{k'}(z,t)} \delta \bar f_{k-k'} (\bm v, z,t) + c.c. \right)\; . \label{eq:elef2} 
\end{eqnarray}
By direct substitution of Eq. (\ref{eq:elef2}) into Eq. (\ref{eq:vlasov}), we obtain
\begin{eqnarray}
& & \left[ 2\Omega + (k+k') v_\parallel - (\omega_k + \omega_{k'}) - i (\partial_t + v_\parallel \partial_z) \right]\delta \bar f_{k+k'}
= \frac{e}{m}  \frac{v_\perp}{4}  \left\{ \delta \bar E_k \left[ \bar \partial_{{\cal E} k}  \right. \right. \nonumber \\
& & \hspace*{1em}  \left. \left. - \frac{1}{v_\perp^2}
\left(1-\frac{k v_{\|}}{\omega_k}\right)  \right] \delta \bar f_{k'}  + \delta \bar E_{k'} \left[ \bar \partial_{{\cal E} {k'}} - \frac{1}{v_\perp^2} 
 \left(1-\frac{k' v_{\|}}{\omega_{k'}}\right) \right]  \delta \bar f_k \right\}, \label{eq:dfk+k'} 
\end{eqnarray}
where we introduced the abbreviated notation 
\begin{eqnarray}
\bar \partial_{{\cal E} k} & \equiv & \left[\frac{k}{\omega_k} \frac{\partial}{\partial v_{\|}}+\left(1-\frac{k v_{\|}}{\omega_k}\right) \frac{1}{v_{\perp}} \frac{\partial}{\partial v_{\perp}}\right]
\nonumber \\ & = & \left[\frac{\partial}{\partial {\cal E}}+\left(1-\frac{k v_{\|}}{\omega_k}\right) \frac{1}{B} \frac{\partial}{\partial \mu}\right] \; ; \label{eq:barp}
\end{eqnarray}
and
\begin{eqnarray}
& & \left[  (k-k') v_\parallel - (\omega_k - \omega_{k'}) - i (\partial_t + v_\parallel \partial_z) \right]\delta \bar f_{k-k'}   = \frac{e}{m}  \frac{v_\perp}{4} 
\left\{ \delta \bar E_k \left[ \bar \partial_{{\cal E} k} + \frac{1}{v_\perp^2} \right. \right.\nonumber \\
& & \hspace*{1em} \times \left.\left.  
\left(1-\frac{k v_{\|}}{\omega_k}\right)  \right] \delta \bar f_{k'}^*  - \delta \bar E_{k'}^* \left[ \bar \partial_{{\cal E} {k'}} + \frac{1}{v_\perp^2} 
 \left(1-\frac{k' v_{\|}}{\omega_{k'}}\right) \right]  \delta \bar f_k \right\}\; .  \label{eq:dfk-k'}
\end{eqnarray}
Note the normalizations on the right hand side of Eqs. (\ref{eq:dfk+k'}) and (\ref{eq:dfk-k'}) to avoid double counting for $k \leftrightarrow k'$; 
and that, on the left hand side, the operators $(\partial_t + v_\parallel \partial_z)$ account for the 
residual slow spatiotemporal dependences consistent with the considered wave packet representation. 
A particular important role is played by the $k=k'$ response in Eq. (\ref{eq:dfk-k'}). It is readily noted that this contribution is not 
characterized by any fast temporal nor spatial dependences. Thus, by all means, $\delta \bar f_0$ (and its c.c.) should be considered as
a modification of the considered original equilibrium due to {\em self interactions} of the fluctuation spectrum; \idest, $f_0$ of Eq. (\ref{eq:f0eq})
and/or Eq. (\ref{eq:f0eqemu}), should be replaced by
\begin{equation}
f_0 \rightarrow f_0 + \sum_k \mathbb R{\rm e} \delta \bar f_0 \; , \label{eq:renorm0}
\end{equation}
with $\delta \bar f_0$ solution of Eq. (\ref{eq:dfk-k'}) for $k' = k$ \cite{Zonca2021}:
\begin{eqnarray}
& & (\partial_t + v_\parallel \partial_z) \delta \bar f_{0}   = \frac{e}{m}  \frac{v_\perp}{4} 
\sum_k i \left\{ \delta \bar E_k \left[ \bar \partial_{{\cal E} k} + \frac{1}{v_\perp^2} \right. \right.\nonumber \\
& & \hspace*{1em} \times \left.\left.  
\left(1-\frac{k v_{\|}}{\omega_k}\right)  \right] \delta \bar f_{k}^*  - \delta \bar E_{k}^* \left[ \bar \partial_{{\cal E} {k}} + \frac{1}{v_\perp^2} 
 \left(1-\frac{k v_{\|}}{\omega_{k}}\right) \right]  \delta \bar f_k \right\}\; .  \label{eq:renorm1}
\end{eqnarray}
If we schematically illustrate Eq. (\ref{eq:dbarfk})
borrowing the Feynman diagram in Fig. \ref{fig:feynman}(a), 
\cite{Zonca2015, Zonca2015b, Chen2016, Zonca2021},
the modification of the original equilibrium defined in Eq. (\ref{eq:renorm0}) corresponds to the simple {\em loop diagram}
in Fig. \ref{fig:feynman}(b); that is, to the {\em emission and reabsorption} of the same-$k$ fluctuations, interpreted as quanta, summed-up over the
whole spectrum.
Quite interesting, the solution of Eq. (\ref{eq:renorm0}), $\delta \bar f_0$, grows secularly in time as $\sim t$ on a 
sufficiently short time scale. Even further, by iterating the procedure and substituting the ``updated'' 
$f_0$, according to Eq. (\ref{eq:renorm0}), into the expression for $\delta \bar f_k$, Eq. (\ref{eq:dbarfk}), we could calculate a correspondingly 
``updated'' expression $\delta \bar f_0$ from Eq. (\ref{eq:renorm1}), whose secularities would grow as $\sim t^2$, and so on. 
This behavior is well known in the general procedure used for deriving kinetic equations in weakly non-ideal systems \cite{vanhove55,prigogine62,balescu63},
and is what dominates the so-called ``principal series'' of secular terms obtained by formal perturbation expansion in powers of fluctuating fields.
If fluctuations of the spectrum are replaced by unstable oscillations, it can be shown that the $\sim t^\ell$ secularities are replaced by terms $\sim (\omega_k/\gamma_k)^\ell$
\cite{montgomery63,Altshul1966}, with $\gamma_k$ the growth rate of the $\omega_k$ fluctuation. The essence, however,
remains unchanged, and the ``principal series'' still dominates the $f_0$ response. In fact, other terms in the perturbation
expansion, at each order, would be at least $\sim 1/(\omega_k \tau_{NL}) \sim (\gamma_k/\omega_k)$ smaller, with $\tau_{NL}$
the characteristic nonlinear time. 
As a matter of fact, the overall response
of the ``principal series'' is obtained from the solution of 
\begin{eqnarray}
& & (\partial_t + v_\parallel \partial_z)  f_{0}   = \frac{e}{m}  \frac{v_\perp}{4} \mathbb R{\rm e}
\sum_k i \left\{ \delta \bar E_k \left[ \bar \partial_{{\cal E} k} + \frac{1}{v_\perp^2} \right. \right.\nonumber \\
& & \hspace*{1em} \times \left.\left.  
\left(1-\frac{k v_{\|}}{\omega_k}\right)  \right] \delta \bar f_{k}^*  - \delta \bar E_{k}^* \left[ \bar \partial_{{\cal E} {k}} + \frac{1}{v_\perp^2} 
 \left(1-\frac{k v_{\|}}{\omega_{k}}\right) \right]  \delta \bar f_k \right\}\; ,  \label{eq:renorm}
\end{eqnarray}
with Eq. (\ref{eq:f0eq}) as initial condition. In fact, it can be verified that the iterative solution of Eqs. (\ref{eq:dbarfk})
and (\ref{eq:renorm}) generates the whole {\em Dyson series} represented by all loop diagrams in Fig. \ref{fig:dyson}.
Thus, we dub Eq. (\ref{eq:renorm}) as {\em Dyson-like} equation \cite{Dyson1949,Schwinger1951,Itzykson80}, noting
that it gives the {\em renormalized electron response} as evolution
of the considered original equilibrium due to self interactions of the fluctuation spectrum \cite{Zonca2015,Zonca2015b,Chen2016,Zonca2021}. 

The Dyson-like equation, Eq. (\ref{eq:renorm}), can be considered as an asymptotic representation of the renormalized
electron response, where the asymptotic expansion parameter is $\sim 1/(\omega_k \tau_{NL}) \sim (\gamma_k/\omega_k) \ll 1$.
Meanwhile, nonlinear interactions in Eqs. (\ref{eq:dfk+k'}) and (\ref{eq:dfk-k'}) generally modify the evolution equation
for $\delta \bar f_k$. Similar to the argument leading us to Eq. (\ref{eq:renorm}), dominant nonlinear interactions are those corresponding 
to emission and reabsorption of the same quanta \cite{vanhove55,prigogine62,balescu63,Altshul1966,Dupree1966,aamodt67,weinstock69,Mima73}.
It can be verified by inspection that these ``diagonal interactions'', which renormalize the evolution operator for $\delta \bar f_k$,
are obtained by two contributions: (i) coupling the $\delta \bar f_{k+k'}$ in Eq. (\ref{eq:dfk+k'}) with the complex conjugate of $k'$ expression
in Eq. (\ref{eq:emforce}); (ii) coupling the $\delta \bar f_{k-k'}$ in Eq. (\ref{eq:dfk-k'}) and its complex conjugate by 
exchanging $k \leftrightarrow k'$ with  the $k'$ expression
in Eq. (\ref{eq:emforce}). The former contribution produces
\begin{eqnarray}
- i \Delta_{1 k} & \equiv & i \frac{e^2}{m^2} \frac{v_\perp^2}{8} \sum_{k'} \delta \bar E_{k'}^* \left[ \bar \partial_{{\cal E} k'} + \frac{2}{v_\perp^2} \left( 1 - \frac{k' v_\parallel}{\omega_{k'}}
\right) \right] \nonumber \\ & &  \times {\cal P}_{k+k'} \delta \bar E_{k'} \left[ \bar \partial_{{\cal E} k'} - \frac{1}{v_\perp^2} \left( 1 - \frac{k' v_\parallel}{\omega_{k'}}
\right) \right] \; , \label{eq:Delta1k}
\end{eqnarray}
to be added on the left hand side in the evolution equation for $\delta \bar f_k$. Here, we have introduced the ${\cal P}_{k+k'}$ {\em propagator}:
\begin{equation}
{\cal P}_{k+k'} \equiv \left[ 2\Omega + (k+k') v_\parallel - (\omega_k + \omega_{k'}) - i (\partial_t + v_\parallel \partial_z) \right]^{-1} \; , \label{eq:Pk+k'}
\end{equation}
defined as the inverse of the operator in the square parentheses.
Meanwhile, the second contribution produces an analogous term
\begin{eqnarray}
- i \Delta_{2 k} & \equiv & i \frac{e^2}{m^2} \frac{v_\perp^2}{8} \sum_{k' \neq k} \delta \bar E_{k'} \bar \partial_{{\cal E} k'} {\cal P}_{k-k'} \delta \bar E_{k'}^* \left[ \bar \partial_{{\cal E} k'} + \frac{1}{v_\perp^2} \left( 1 - \frac{k' v_\parallel}{\omega_{k'}}
\right) \right] \; , \label{eq:Delta2k}
\end{eqnarray}
with 
\begin{equation}
{\cal P}_{k-k'} \equiv \left[  (k-k') v_\parallel - (\omega_k - \omega_{k'}) - i (\partial_t + v_\parallel \partial_z) \right]^{-1} \; . \label{eq:Pk-k'}
\end{equation}
Thus, taking diagonal interactions into account, Eq. (\ref{eq:dbarfk}) is replaced by
\begin{eqnarray}
\delta \bar{f}_k & = & \frac{e}{m} {\cal P}_k v_\perp \delta \bar E_k \bar \partial_{{\cal E} k}  f_0 \; , \label{eq:dbarfkrenorm} \\
{\cal P}_k & \equiv & \left[  \Omega  + k v_\parallel - \omega_k - i (\partial_t + v_\parallel \partial_z) - \Delta_k \right]^{-1} \; ; \label{eq:Pk}
\end{eqnarray}
with $\Delta_k \equiv \Delta_{1k} + \Delta_{2k}$ describing both nonlinear resonance frequency shift as well as {\em resonance broadening}
\cite{Dupree1966,aamodt67,weinstock69,Mima73}. In the linear limit, Eq. (\ref{eq:dbarfkrenorm}) readily recovers Eq. (\ref{eq:dbarfk}).
Meanwhile, allowing for leading order nonlinear interactions, Eqs. (\ref{eq:renorm}) and (\ref{eq:dbarfkrenorm}) fully describe the
renormalized energetic electron response, self-consistently evolving with the fluctuation spectrum, described by Eqs. (\ref{eq:actionevolve}) and (\ref{eq:phaseevolve}).
In our applications to chorus chirping in Sec. \ref{sec:chorus}, resonance broadening will be dropped in Eq. (\ref{eq:Pk}), since its
effect is typically small in the parameter range where chorus is usually observed (cf. Sec. \ref{sec:conserve}).
As anticipated in Sec. \ref{sec:choruslin}, Eq. (\ref{eq:WGamma4lin}) for the wave particle power exchange remains valid nonlinearly, provided the
renormalized expression of the propagator is used as in Eq. (\ref{eq:Pk}), instead of the linearized (algebraic) expression, $\left[  \Omega  + k v_\parallel - \omega_k\right]^{-1}$.

\subsection{Conservation properties and phase space transport}
\label{sec:conserve}

Let's reconsider the Dyson-like equation, Eq. (\ref{eq:renorm}), and rewrite it as 
\begin{equation}
(\partial_t + v_\parallel \partial_z)  f_{0}   + \frac{1}{v_\perp} \frac{\partial}{\partial v_\perp} \left( v_\perp S_\perp \right)
 + \frac{\partial}{\partial v_\parallel} S_\parallel  = 0 \; . \label{eq:renormcon}
\end{equation}
Here, we have introduced
\begin{eqnarray}
S_\perp & \equiv & \sum_k \frac{i}{4} \left( 1 - \frac{k v_\parallel}{\omega_k} \right) \delta \bar E_k^* \delta \bar f_k  + c.c. \; , \label{eq:Sperp} \\
S_\parallel & \equiv & \sum_k \frac{i}{4} \frac{k v_\perp}{\omega_k}  \delta \bar E_k^* \delta \bar f_k  + c.c.  \; . \label{eq:Sparallel}
\end{eqnarray}
Integrating Eq. (\ref{eq:renormcon}) in velocity space, $\left \langle ... \right\rangle = 2\pi \iint v_\perp dv_\perp d v_\parallel (...)$ with the notation 
introduced in Sec. \ref{sec:choruslin},  we have the continuity equation
\begin{equation}
\partial_t n + \partial_z (n u) = 0 
 \; , \label{eq:continuity}
\end{equation}
where $nu$ is the parallel electron flux. Thus, particles are conserved. Similarly, taking the product of 
Eq. (\ref{eq:renormcon}) times $m v_\parallel$ and  performing the same integration, we have
\begin{equation}
\partial_t ( mnu )+ \partial_z \left \langle m v_\parallel^2 f_0 \right\rangle = \left\langle m S_\parallel \right\rangle 
 \; . \label{eq:momentum}
\end{equation}
Equation (\ref{eq:momentum}), expresses parallel momentum conservation accounting for wave-particle momentum exchange. Finally,
multiplying Eq. (\ref{eq:renormcon}) by $m{\cal E}$ and integrating in velocity space, denoting particle kinetic energy density by 
${\cal K} = \left\langle m {\cal E} f_0 \right\rangle$, we have
\begin{eqnarray}
\partial_t {\cal K} + \partial_z \left \langle m v_\parallel {\cal E} f_0 \right\rangle & = & \left\langle m \left( v_\perp S_\perp + v_\parallel S_\parallel \right) \right\rangle 
= -  \sum_k \mathbb I{\rm m} \left\langle e \frac{v_\perp}{2}  \delta \bar E_k^* \delta \bar f_k  \right\rangle \nonumber \\
& = &  
 - \sum_\omega \Gamma (z,t,\omega) \omega \frac{\partial D_w}{\partial \omega} \left. \frac{|\delta \bar{\bm E}_{\perp k}|^2}{8 \pi} \right|_{k = K(z,\omega)} 
\nonumber \\ & = &  - \left[ \partial_t \sum_k W_{wk} + \partial_z \sum_k ( v_{gk}  W_{wk} ) \right]
 \; . \label{eq:energy}
\end{eqnarray}
Here, in the second line we have used Eq. (\ref{eq:WGamma}) and, in the third line, we have rewritten Eq. (\ref{eq:actionevolve}) introducing
the wave energy density in the $k$ fluctuation as $W_{w k} \equiv \omega_k \partial_{\omega_k} D_w |\delta \bar{\bm E}_{\perp k}|^2/(16\pi)$ \cite{Bernstein75}.
Thus, Eq. (\ref{eq:energy}) expresses energy conservation. In conclusion, the Dyson-like equation, Eq. (\ref{eq:renorm}), properly
describes phase space transport with all the necessary conservation properties.

A variety of different nonlinear dynamics behaviors are described by the self-consistent equations for fluctuation spectrum, Eqs. (\ref{eq:actionevolve})
and (\ref{eq:phaseevolve}), and 
supra-thermal electron response, Eqs. (\ref{eq:renorm}) and (\ref{eq:dbarfkrenorm}). Starting from very low fluctuation level, \exgra, with whistler waves excited by wave particle interactions out of thermal noise,
the spectrum is generally very broad, with width $\Delta \omega \sim \omega$, as noted in Sec. \ref{sec:choruslin}. 
The resonance broadening effect is typically small, $\sim \omega_{{\rm tr} k}^4/(\gamma_k \Delta \omega^3)$ and can be neglected. Meanwhile, the dense whistler wave spectrum causes particles to undergo {\em quasilinear}-type diffusion, consistently described by Eq. (\ref{eq:renorm}), which reduces to the quasilinear
diffusion equation in this limit \cite{galeev65,Altshul1966}. The diffusion time is $\tau_{\rm diff} \sim \Delta \omega^3/\omega_{{\rm tr} k}^4 \gg 1/\Delta \omega$, consistent
with having small {\em Kubo number}, $K \sim \omega_{{\rm tr} k}/\Delta \omega \ll 1$. Since $\gamma_k \tau_{\rm diff} 
\sim \gamma_k \Delta \omega^3/\omega_{{\rm tr} k}^4 \gg 1$,
the wave can be significantly amplified before any significant diffusion has occurred. However, due to the convective nature of the whistler instability,
saturation of  fluctuation growth can be reached because the wave packet moves out of the source region and before nonlinear behavior
becomes noticeable. Thus, we expect that there is a threshold in driving strength, due to a combination of initial level of whistler waves and convective amplification
by linear instability, which determines the observed nonlinear behavior (cf. Sec. \ref{sec:chorus}). Optimal ordering for the fluctuation amplitude needed to observe
nonlinear oscillations would be obtained for 
\begin{equation}
\tau_{NL}^{-1} \sim \gamma_k \sim \omega_{{\rm tr} k} \; . \label{eq:optimal}
\end{equation}
When this condition is met, the mode does not stop growing by the well-known wave particle trapping mechanism discussed by \cite{ONeil1968,ONeil1971} for the
beam-plasma instability, since phase mixing is slowed down by the presence of a broad spectrum (cf. discussion above). 
Estimating ${\cal P}_k \sim \tau_{NL}$ in Eq. (\ref{eq:Pk}),  and $\bar \partial_{{\cal E} k} \sim (k^2/\omega_k)/(k\Delta v)$, with $k \Delta v$ the resonance width,
Eq. (\ref{eq:renorm}) suggests that nonlinear whistler wave packet dynamics be characterized by
\begin{equation}
\tau_{NL}^{-1} \sim  \frac{\omega_{{\rm tr} k}^2}{k\Delta v} \gaeq k\Delta v \; ; \;\;\;\;\; \Leftrightarrow \;\;\;\;\; \tau_{NL}^{-1} \gaeq  \omega_{{\rm tr} k} \gaeq k\Delta v \; . \label{eq:optimal2}
\end{equation}
Note that this condition implies the Kubo number be of $\sim {\cal O}(1)$ and the corresponding nonlinear behavior described by Eq. (\ref{eq:renorm})
be non-perturbative. This is the reason why the Dyson-like equation must be solved as is and not by iterative perturbation expansion, since the 
Dyson series terms would have the  Kubo number as expansion parameter.
In this regime, that is typical of chorus excitation (cf. Sec. \ref{sec:chorus}), resonance broadening is still negligible provided that the system is not too
strongly driven. In fact, estimating ${\cal P}_{k\pm k'} \sim \Delta \omega^{-1}$ in Eqs. (\ref{eq:Pk+k'}) and (\ref{eq:Pk-k'}), Eqs. (\ref{eq:Delta1k}) and 
 (\ref{eq:Delta2k}) yield
\begin{equation}
\Delta_k \sim \frac{1}{\tau_{NL} \Delta \omega} \frac{1}{\tau_{NL}} <  \frac{1}{\tau_{NL}} \; . \label{eq:optimal3}
\end{equation}
For the quasi-coherent chorus spectrum $\Delta \omega/\omega_{{\rm tr} k} \gaeq \tau_{NL} \Delta \omega > 1$, and Eq. (\ref{eq:optimal3}) is verified. 
Increasing the driving strength makes $\Delta \omega$ smaller. Eventually, when $\tau_{NL} \Delta \omega \sim 1$, a strong effect is expected from 
resonance broadening (cf., \exgra, \cite{laval84,laval99}) and the coherent nature of chorus excitation will be lost. Unlike the lower threshold in driving
strength for chorus to be triggered, which is sharp since nonlinear dynamics must set-in, the higher threshold for resonance broadening to smear out
chorus is expected to be a smooth transition where coherent behavior is gradually lost.

\section{Chorus emission and frequency chirping}
\label{sec:chorus}

As application of the theoretical framework presented in Sec. \ref{sec:choruslin} and Sec. \ref{sec:phasespace},
we address the chorus emission in the Earth's magnetosphere as manifestation of nonlinear dynamics of whistler
waves excited by supra-thermal electrons. We first give a brief summary of chorus observations and discuss
their qualitative features in Sec. \ref{sec:chorusobs}, based on the present theoretical approach 
and corresponding understandings. For application purposes, in Sec. \ref{sec:reddyson_der} 
we then develop a reduced Dyson model for chorus emission. The reduced description is based on
two simplifying assumptions \cite{Zonca2021}: (i) considering  
a non-uniform source of hot electrons, localized about
the equator, neglecting magnetic field non-uniformity (cf. Sec. \ref{sec:whistlerinst}); and (ii)
constructing simplified nonlinear expressions for the $W(z,t,\omega)$ and $\Gamma(z,t,\omega)$
functions in Eq. (\ref{eq:WGamma4lin}) rather than solving for the whole phase space renormalized
electron response. 
These expressions, given by Eqs. (\ref{eq:redWbar0}) and (\ref{eq:redGammabar0}), 
are systematically derived from Eq. (\ref{eq:WGamma4lin}), with its nonlinear extension Eq. (\ref{eq:WGammaNL}),
and Eq. (\ref{eq:renorm}) in Ref. \cite{Zonca2021}, where interested readers can find all technical details. Here, derivation
is only sketched to focus on the underlying physics rather than on the mathematical aspects.
One important result of the reduced Dyson model is that it can
analytically demonstrate \cite{Zonca2021}  the chorus chirping rate as conjectured by
\cite{Vomvoridis1982}. Furthermore, it draws fundamental analogies 
between chorus and frequency chirping phenomena observed in magnetic confinement fusion plasmas,
which will be further discussed in Sec. \ref{sec:conclusions}.
Numerical solutions of this reduced model are presented in Sec.
\ref{sec:reddyson_sol}, confirming the behaviors anticipated in Sec. \ref{sec:conserve}
and Sec. \ref{sec:chorusobs}. These results will then be used is Sec. \ref{sec:compare}
to qualitatively and quantitatively compare the present analysis of chorus emission 
and frequency chirping with other existing models.

\subsection{Chorus observation and qualitative features}
\label{sec:chorusobs}

\begin{figure}
  \centering
  \includegraphics[width=0.85\textwidth]{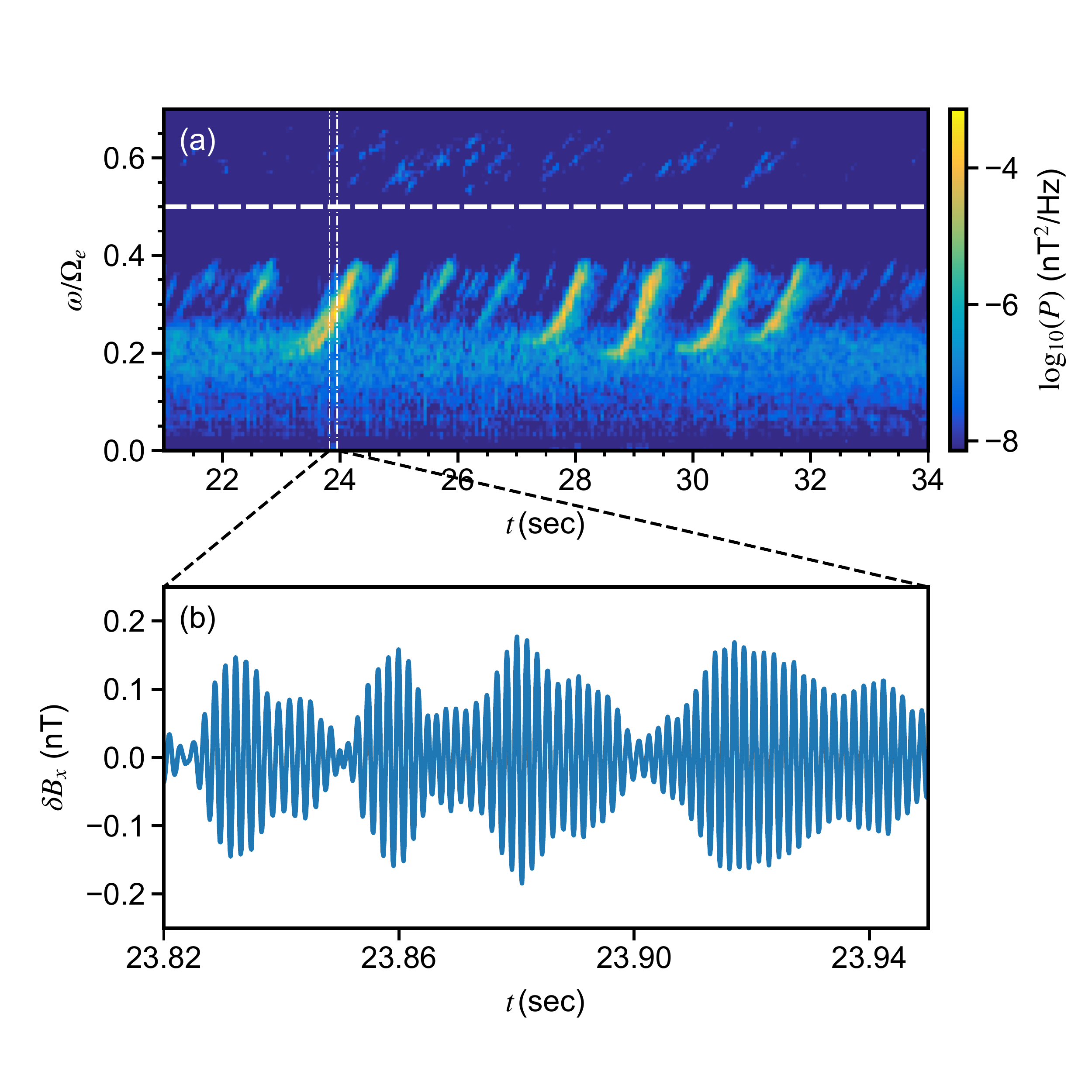}
  \caption{(a) THEMIS observation of rising-tone chorus waves on July 26, 2008 at $L=7.6$, magnetic local time MLT =11.7\,h, and magnetic latitude MLAT=2.7$^\circ$. The x-axis is the seconds since 13:26:00\,UTC, and colorcoded is the logarithm of wave magnetic field power spectral density. The white dashed line indicates half electron cyclotron frequency and the dash-dotted lines indicate the starting and ending time of the waveform ($\delta B_x$) plotted in (b), showing quasi-coherent nature of chorus. Here $\delta B_x$ is one of the magnetic field component perpendicular to the local background magnetic field.}
  \label{fig:themis}
\end{figure}

Chorus waves, illustrated in Figure \ref{fig:themis}, are an important type of electromagnetic whistler mode waves frequently observed in various planetary magnetospheres \cite{Tsurutani1974,Burtis1976, Hospodarsky2008,Menietti2008c}. These waves have been demonstrated to play key roles in radiation belt electron acceleration \cite{Horne1998,Horne2005b,Thorne2010a,Thorne2013b,Reeves2013}, and precipitation of energetic electrons into the atmosphere to form diffuse \cite{Thorne2010} and pulsating aurora \cite{Nishimura2010}. Chorus waves have attracted significant research interest also because of their special properties and, thus, their generation mechanism. Observationally, the spectrograms of chorus waves consist of a series of quasi-coherent narrowband discrete elements with frequency chirping. The chirping rate can be either positive or negative, leading to the so-called rising-tone or falling-tone chorus, respectively, although more complicated spectral shapes have also been occasionally reported \cite{Burtis1976}. For rising-tone chorus in the terrestrial magnetosphere, each element lasts about $\mathcal{O}(10^{-1})$\,s \cite{Teng2017}, and the frequency range of an element is on the order of $0.1\Omega_{e}$ or $\mathcal{O}(10^2)$\,Hz \cite{Burtis1976}; therefore, the typical frequency chirping rate is about kHz/s. Observations have further determined that the source region of chorus waves is about $3^\circ$ in latitude near the equator \cite{Santolik2003,Teng2018}. Correspondingly, the fast frequency chirping of chorus cannot be a result of wave propagation and dispersion as in the case of lighting-generated whistlers. The current consensus is that the chirping of chorus is a direct result of coherent nonlinear wave particle interactions with energetic electrons. What is under debate is how the nonlinear interactions lead to chirping and how to properly describe the process theoretically \cite{Nunn1974,Vomvoridis1982,Omura2008,Zonca2017,Tao2020}.    

Although observations report both falling- and rising-tone chorus\footnote{We will come back on this issue later in this section.},
there is a clear prevalence of rising-tone chorus. In Sec. \ref{sec:singlep}, we showed that ``phase locked'' resonant
particles, which are deeply trapped near the O-points of the phase space holes formed by the frequency chirping whistler wave packet
and shown in Fig. \ref{fig:wavetrap}, actually gain energy and magnetic moment. Thus, they are incompatible with the wave packet
growth. To better clarify this point, let us reconsider Fig. \ref{fig:wavetrap} and note that, for a rising-tone chorus, the phase space structures
O-points are moving toward the lower parallel energies. Meanwhile, particles on open trajectories are moving from positive $\dot \phi$ on the
top to negative $\dot \phi$ on the bottom, thereby gaining parallel energy but loosing magnetic moment and total energy as a whole.
These particles are more numerous and most efficiently contribute to the power exchange and mode growth, consistent with the
clear picture of resonant wave particle power exchange recently discussed by \cite{escande18} (``...synchronization of almost 
resonant passing particles...''). 
The major importance of resonant passing electrons for whistler wave growth has been pointed out by \cite{Karpman1974,Shklyar2009}
and by \cite{Shklyar2011} for electron acceleration by whistler waves.
This qualitatively explains the
growth mechanism of a rising-tone chorus wave packet and is clearly connected with {\em phase bunching}, since the
maximum power transfer from particles to waves is occurring in a relatively narrow region of the wave-particle phase
$\phi$: in Fig. \ref{fig:wavetrap}, for particles moving from positive to negative $\dot \phi$ at $\phi/\pi \sim -0.5, 1.5 ...$ .

\begin{figure}[t]
	\begin{center}
		\resizebox{0.6\textwidth}{!}{\includegraphics{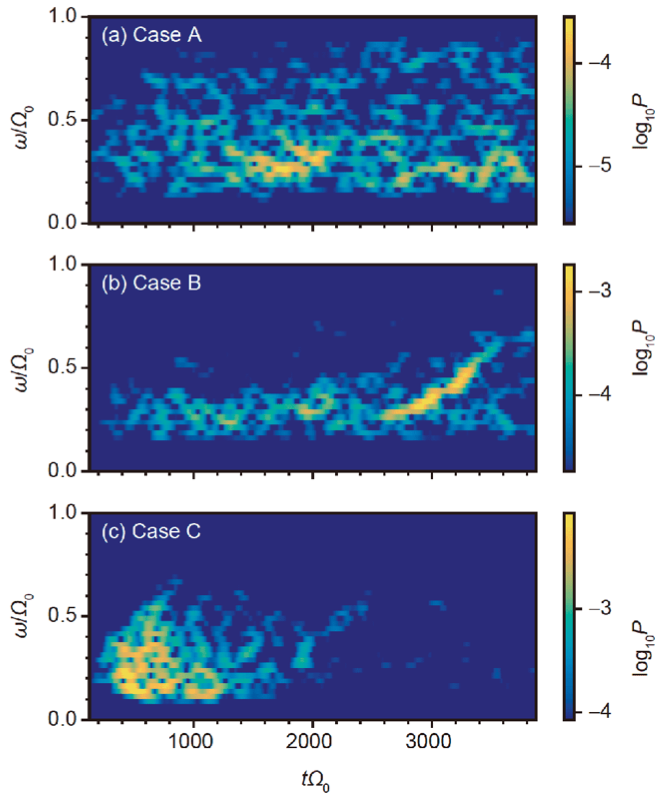}}
	\end{center}
\caption{Contour plot of the whistler wave power spectrum (color bar) obtained by DAWN code \cite{Tao2014b}
PIC simulations illustrating 
different qualitative behavior for increasing supra-thermal electron drive (From original figure in Ref. \cite{Tao2020}). 
Case (A), (B) and (C), are characterized by the 
same parameters as in Fig. \ref{fig:WGamma} except for $w_{\perp e} = 0.32c$ (a),  $w_{\perp e} = 0.35c$ (b)
and $w_{\perp e} = 0.6c$ (c). Here, the original notation $\Omega_0$ stands for $\Omega_e$.}
\label{fig:whistler_cases}
\end{figure}

\begin{figure}[t]
	\begin{center}
		\resizebox{0.6\textwidth}{!}{\includegraphics{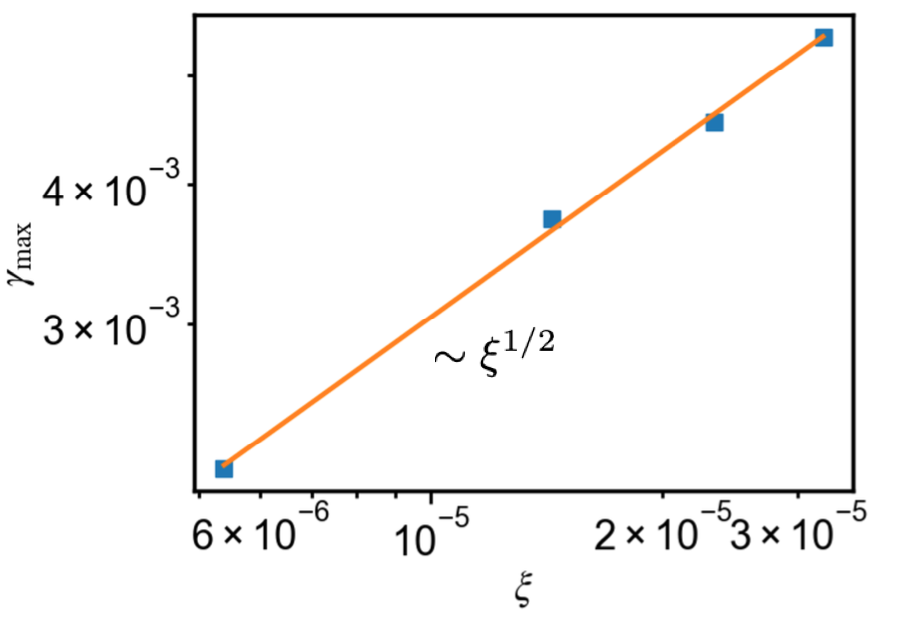}}
	\end{center}
\caption{Scaling of the threshold in peak of the linear drive, $\gamma_{\rm max}$, with the magnetic field nonuniformity
parameter, $\xi$, needed for chorus emission in DAWN code PIC simulations \cite{Tao2014b}.}
\label{fig:thresh1}
\end{figure}

Recalling Sec. \ref{sec:conserve} concluding discussion on the variety of different nonlinear dynamics behaviors 
that are described by the present theoretical framework, one crucial parameter is the driving strength of the whistler wave
spectrum. In particular, two thresholds
are expected: (i) a lower threshold for chorus emission, expected to be quite sharp; (ii) and an upper threshold or, more properly,
smooth transition to a regime where resonance broadening dominates and destroys the chorus coherence. This qualitative
behavior is illustrated by the particle in cell (PIC) numerical simulation results by the DAWN code \cite{Tao2014b}
in Fig. \ref{fig:whistler_cases}, where the linear drive is increased from case A to case C.
The first transition, according to the argument provided above and in Sec. \ref{sec:conserve} should be connected with
the strength of linear drive; that is with the peak of the linear drive at $z=0$; \idest, $\gamma_{\rm max}$. In fact, 
noting Eq. (\ref{eq:Ikevolve}) and neglecting, for simplicity, magnetic field non-uniformity vs. the non-uniformity of the 
energetic electron source (cf. Sec. \ref{sec:choruslin}), we have that the maximum amplification of the linear wave packet
is \cite{Zonca2017}
\begin{eqnarray}
I/I_{\omega 0} & = & \exp \left( 2 \int_{-\infty}^\infty \frac{\Gamma(z,0,\omega)}{v_g(z,\omega)} dz \right) \label{eq:drate} \\
& \simeq & \exp \left( 2 \frac{\gamma_{\rm max}}{v_g(0,\omega)} \int_{-\infty}^\infty \zeta^4 dz \right) = \exp \left( 2 \frac{\gamma_{\rm max}}{v_g(0,\omega)} \frac{\pi}{(A \xi)^{1/2}} \right) \nonumber \; .
\end{eqnarray}
This suggest that $\gamma_{\rm max}$ should scale with $\xi^{1/2}$ to have the required minimum driving rate 
for successfully triggering chorus emission, as shown in Fig. \ref{fig:thresh1} using DAWN code simulation results \cite{Tao2014b}.

\subsection{Reduced Dyson model for chorus emission: derivation}
\label{sec:reddyson_der}

The derivation of the reduced Dyson model for chorus emission is originally given in Ref. \cite{Zonca2021}.
It is beyond the intended scope of the present tutorial manuscript to enter into the same detailed analysis. We
rather give a sketch of that derivation, which we try to make accessible to a broad readership,
leaving interested readers to the original reference for details.
As anticipated in the introductory remarks of Sec. \ref{sec:chorus}, we can have a first simplification
noting the $\sim \zeta^4 \sim (1 + A \xi z^2)^{-2}$ scaling of Eq. (\ref{eq:WGammalin}), when
the magnetic field non-uniformity is neglected; that is, when $\Omega \simeq \Omega_e$ and 
$\xi z^2 \ll 1$ are assumed. This allows us to rewrite Eq. (\ref{eq:WGamma4lin}) as 
\begin{eqnarray}
& & W(z, t, \omega)+i \Gamma(z, t, \omega)  =  \nonumber \\
& & \hspace*{2em} = \frac{\omega (\Omega_e - \omega)^2}{\Omega_e} \frac{n_e}{n} \zeta^4 \left\langle \frac{v_\perp^2}{2} {\cal P}_k
\left[ \frac{k v_\perp^2}{2 \omega} \frac{\partial \hat f_0}{\partial v_{ \|}} - \frac{\Omega_e}{\omega} \hat f_0 \right]_{k = K(z=0,\omega)} \right\rangle
 \; , \label{eq:WGammaNL}
\end{eqnarray}
where we let $(1 - k v_\parallel/\omega) \simeq \Omega_e/\omega$ for nearly resonant particles and $\hat f_0 = f_0/n_e$ 
represents the normalized electron response computed for a uniform anisotropic plasma. Furthermore,  all
whistler wave dispersion relation and dispersive properties, including the ${\cal P}_k$ propagator defined in Eq. (\ref{eq:Pk}), 
are intended to be computed at $k=K(z=0,\omega)$ (cf. Sec. \ref{sec:choruslin}) \cite{Zonca2021}.
For example, the integral operator defined in Eq. (\ref{eq:Tktrans}) becomes algebraic, $T_\omega (z) = z/v_{g\omega}$,
and its inverse is also simplified, $T^{-1}_\omega (t) = v_{g\omega} t$, with $v_{g\omega} = v_g(z=0,\omega)$.
Note that Eq. (\ref{eq:WGammaNL}) is fully nonlinear, although the linear limit is readily recovered for 
${\cal P}_k = [\Omega_e + k v_\parallel - \omega]^{-1}$ and $\hat f_0$ given by the properly rescaled
bi-Maxwellian in Eq. (\ref{eq:f0eq}). Following Ref. \cite{Zonca2021}, it is also useful to 
 introduce the rescaled phase shift and driving rate,
$\bar W(z, t, \omega)$ and $\bar \Gamma(z, t, \omega)$, defined as
\begin{equation}
W(z, t, \omega)+i \Gamma(z, t, \omega) \equiv \zeta^4 \frac{\omega(\Omega_e-\omega)^2}{\Omega_e^2 (1-v_{r\omega}/v_{g\omega})^2}
 \left[ \bar W(z, t, \omega)+i \bar \Gamma(z, t, \omega) \right]
 \; , \label{eq:scalWGamma}
\end{equation}
where $v_{r\omega} = v_r(z=0,\omega)$. As a result, Eq. (\ref{eq:WGammaNL})
can be rewritten as
\begin{equation}
\bar W(z, t, \omega)+i \bar \Gamma(z, t, \omega) = \frac{n_e}{n} \left( 1 - \frac{v_{r\omega}}{v_{g\omega}} \right)^2
\left\langle \Omega_e {\cal P}_k
\left[ \frac{k v_\perp^2}{2 \omega} \frac{\partial}{\partial v_\parallel} -  \frac{\Omega_e}{\omega} \right] \hat f_0 \right\rangle \; . \label{eq:WGammabarNL}
\end{equation}
Equation (\ref{eq:WGammabarNL}) hints at the reason underlying the construction of the reduced Dyson model; 
that is, we could readily solve the equations for the fluctuation spectrum, Eqs. (\ref{eq:actionevolve}) and (\ref{eq:phaseevolve}), 
if we could determine $\bar W$ and
$\bar \Gamma$ without solving for the whole phase space renormalized
electron response, $\hat f_0$. 
This is the second simplification alluded at in the introduction to Sec. \ref{sec:chorus}.

In order to make further analytic progress, let us consider $\hat f_0 \equiv (\partial_t + v_\parallel \partial_z)^{-1} (\partial_t + v_\parallel \partial_z) \hat f_0$
on the right hand side of Eq. (\ref{eq:WGammabarNL}), and use Eq. (\ref{eq:renorm}) to rewrite $(\partial_t + v_\parallel \partial_z) \hat f_0$
therein. The advantage of this formal manipulation will be clear later. We can also consider the nonlinear evolution of 
the whistler wave spectrum at one fixed $z$, chosen such that $z$ is outside the ``hot'' electron source region and, consistent with
Eqs. (\ref{eq:Ikevolve}) and (\ref{eq:phikevolve}) as well as Fig. \ref{fig:Ivarphi_omega0}, the 
fluctuation spectrum depends only on $t - z/v_{g\omega}$. In this way, we can focus on the nonlinear
behaviors, separating them from those due to linear wave packet propagation. As a consequence, the same kind of dependence is
reflected upon $\hat f_0$ and $\delta \hat{\bar f}_k \equiv \zeta^{-4} \delta \bar  f_k/n_e$; that is, the rescaled supra-thermal
electron response consistent with the earlier definition of $\hat f_0$. Note that, similar to $\hat f_0$,
also $\delta \hat{\bar f}_k$ corresponds to electron response in uniform plasma, all spatial dependences
being accounted for by the $\zeta^4$ factor in Eq. (\ref{eq:scalWGamma}).
Meanwhile, this simplified ``local'' limit is readily obtained from the more general case by  
replacing $(\partial_t + v_\parallel \partial_z) \rightarrow (1-v_{r\omega}/v_{g\omega}) \partial_t$ 
for nearly resonant particles \cite{Zonca2021}.

Based on these assumptions, on the right hand side of Eq. (\ref{eq:renorm}) we can write
\begin{eqnarray}
& & i \left( \delta \bar E_k {\cal P}_k^* \delta \bar E_k^* -  \delta \bar E_k^* {\cal P}_k \delta \bar E_k \right) \nonumber \\
& & \hspace*{2em} =
i \left( \delta \bar E_k {\cal P}_k^* {\cal P}_k {\cal P}_k^{-1} \delta \bar E_k^* -  \delta \bar E_k^* {\cal P}_k {\cal P}_k^* {\cal P}_k^{*-1} \delta \bar E_k \right) 
\nonumber \\
& & \hspace*{2em} \simeq
2 |\delta \bar E_k| |{\cal P}_k|^2 (1-v_{r\omega}/v_{g\omega}) \partial_t |\delta \bar E_k| \; , \label{eq:integrop1} 
\end{eqnarray}
where, consistent with the optimal ordering for chorus excitation, Eq. (\ref{eq:optimal2}), discussed in Sec. \ref{sec:conserve},
we dropped resonance broadening \cite{Zonca2021}; and we introduced the notation
\begin{equation}
{\cal P}_k^* {\cal P}_k \simeq {\cal P}_k {\cal P}_k^* \simeq |{\cal P}_k|^2 = \left[ (\Omega_e + k v_\parallel - \omega)^2 + 
(1-v_{r\omega}/v_{g\omega})^2 \partial_t^2 \right]^{-1} \; . \label{eq:Pk2}
\end{equation}
Note that the operators action on the phase dependences in $\delta \bar E_k$ and its complex conjugate cancel out
and only $\partial_t |\delta \bar E_k|$ survives. With help of these identities, and substituting Eq. (\ref{eq:renorm}) on
the right hand side of Eq. (\ref{eq:WGammabarNL}), that equation can be cast as
\begin{eqnarray}
\bar W (\omega) +i \bar \Gamma (\omega) & =  & \frac{n_e}{n} \left( 1 - \frac{v_{r\omega}}{v_{g\omega}} \right)^2
\left\langle \frac{v_\perp^4}{2} \Omega_e {\cal P}_k   \left( \frac{k}{\omega} \frac{\partial}{\partial v_\parallel} - \frac{2}{\left\langle v_\perp^2\right\rangle} \frac{\Omega_e}{\omega} \right)  \right. \nonumber \\
& & \times  (1-v_{r\omega}/v_{g\omega})^{-1} \partial_t^{-1} \sum_{k'}  \frac{e^2}{2 m^2} |\delta \bar E_{k'}| \frac{k'}{\omega'} \frac{\partial}{\partial v_\parallel} |{\cal P}_{k'}|^2 \nonumber \\
& &  \times  (1-v_{r\omega'}/v_{g\omega'}) \left. \partial_t |\delta \bar E_{k'}| \left( \frac{k'}{\omega'} \frac{\partial}{\partial v_\parallel} - \frac{2}{\left\langle v_\perp^2\right\rangle} \frac{\Omega_e}{\omega'} \right) \hat f_0 \right\rangle . \label{eq:redWGammabar0}
\end{eqnarray}
Here, we have explicitly denoted only the frequency dependence of $\bar W$ and $\bar \Gamma$, 
and left implicit the dependence on $t - z/v_{g\omega}$. Furthermore, we have introduced the notation $\left\langle v_\perp^2 \right\rangle \equiv 
\left\langle v_\perp^2 \hat f_0 \right\rangle/\left\langle \hat f_0 \right\rangle$.
Interested readers can find in Ref. \cite{Zonca2021} all technical details involved in the derivation of Eq. (\ref{eq:redWGammabar0})
from Eqs. (\ref{eq:renorm}) and (\ref{eq:WGammabarNL}). Velocity space integrals in Eq. (\ref{eq:redWGammabar0}) can be carried
out assuming that the fluctuation spectrum is quasi-coherent (narrow) and representing operators in the Laplace space (cf., \exgra, 
\cite{Zonca2015b,Zonca2015,Chen2016,Tao2020}). Separating real and imaginary parts, and noting that, for nearly resonant particles,
\begin{equation}
\frac{1}{k} \frac{\partial}{\partial v_\parallel}\simeq\frac{1}{k} \frac{\partial}{\partial v_{r\omega}} = \frac{1}{(1-v_{r\omega}/v_{g\omega})} \frac{\partial}{\partial \omega} \; ,
\label{eq:urel4}
\end{equation}
one can finally show, after lengthy but
straightforward algebra \cite{Zonca2021}
\begin{eqnarray}
\Omega_e^{-1} \partial_t \bar W (\omega) & =  & \left[ \Omega_e \frac{\partial}{\partial \omega} - \frac{2 \Omega_e^2}{k^2 \left\langle v_\perp^2\right\rangle} 
\left( 1 - \frac{v_{r\omega}}{v_{g\omega}} \right) \right] \Omega_e \frac{\partial}{\partial \omega} \nonumber \\
& & \times \sum_{k'} \frac{\left( \omega' - \omega \right)}{2 \Omega_e} \left[ \frac{\left( \omega' - \omega \right)^2}{4 \Omega_e^2}+\Omega_e^{-2} \partial_t^2\right]^{-1} \nonumber \\
& &  \times \frac{\left\langle\left\langle \omega_{{\rm tr}k'}^4 \right\rangle\right\rangle}{4 \Omega_e^4 (1-v_{r\omega'}/v_{g\omega'})^4}
\left( \frac{\bar \Gamma (\omega) + \bar \Gamma (\omega')}{2} \right) \; , \label{eq:redWbar0}
\end{eqnarray}
where $\left\langle\left\langle \omega_{{\rm tr}k}^4 \right\rangle\right\rangle = \left\langle v_\perp^2 \omega_{{\rm tr}k}^4 \hat f_0\right\rangle
/\left\langle v_\perp^2 \hat f_0 \right\rangle$ is the velocity space averaged wave particle trapping frequency; and, defining $\bar \Gamma_{NL} = \bar \Gamma - \bar \Gamma_L$, with $\bar \Gamma_{L}$ the  linear (initial) normalized hot electron driving rate, 
\begin{eqnarray}
\bar \Gamma_{NL} (\omega) & =  & \left[ \Omega_e \frac{\partial}{\partial \omega} - \frac{2 \Omega_e^2}{k^2 \left\langle v_\perp^2\right\rangle} 
\left( 1 - \frac{v_{r\omega}}{v_{g\omega}} \right) \right] \Omega_e \frac{\partial}{\partial \omega} \nonumber \\
& & \times \sum_{k'} \left[ \frac{\left( \omega' - \omega \right)^2}{4 \Omega_e^2}+\Omega_e^{-2} \partial_t^2\right]^{-1}  
\nonumber \\ & &  
\times \frac{\left\langle\left\langle \omega_{{\rm tr}k'}^4 \right\rangle\right\rangle}{4 \Omega_e^4 (1-v_{r\omega'}/v_{g\omega'})^4}
\left( \frac{\bar \Gamma (\omega) + \bar \Gamma (\omega')}{2} \right) \; . \label{eq:redGammabar0}
\end{eqnarray}
Equations (\ref{eq:redWbar0}) and (\ref{eq:redGammabar0}) have been significantly simplified with 
respect to the original Dyson-like equation, Eq. (\ref{eq:renorm}). Together with Eqs. (\ref{eq:Ikevolve}) and (\ref{eq:phikevolve}),
they provide the {\em reduced Dyson
model} equations for chorus nonlinear dynamics and the most important novel result
of \cite{Zonca2021}. Despite they are complicated nonlinear integro-differential equations, they allow
us to understand the mechanism underlying chorus chirping. In fact, noting the optimal ordering of Eq. (\ref{eq:optimal2})
introduced in Sec. \ref{sec:conserve}, we expect that nonlinear behavior in Eqs. (\ref{eq:redWbar0}) and (\ref{eq:redGammabar0}) 
may change when, in the first square parenthesis on the right hand side, $\Omega_e \partial_\omega$ dominates and,
in the second square parenthesis, $(\omega' - \omega)^2$ can be dropped with respect to $4\partial_t^2$. As a consequence
of this and of the fact that gradients in the $\omega$ space are dominated by the nonlinear behavior, Eq. (\ref{eq:redGammabar0}) 
becomes \cite{Zonca2021}
\begin{equation}
\frac{\partial^{2}}{\partial t^{2}} \bar{\Gamma}_{N L}(\omega) \simeq\left(\sum_{k'} \frac{\left\langle\left\langle\omega_{\mathrm{tr} k'}^{4}\right\rangle\right\rangle}{4\left(1-v_{r \omega'} / v_{g \omega'}\right)^{4}}\right) \frac{\partial^{2}}{\partial \omega^{2}} \bar{\Gamma}_{N L}(\omega)
\; . \label{eq:ballistic}
\end{equation}
Equation (\ref{eq:ballistic}) has a ``self-similar solution that {\em ballistically propagates} in $\omega$-space at a rate given by the square root of the quantity in parentheses on the right hand side'' \cite{Zonca2021}. Correspondingly, structures in the supra-thermal electron phase space also propagate at the same rate
as described by the Dyson-like equation, Eq. (\ref{eq:renorm}) \cite{Chen2016,Zonca2015b,Zonca2015}. This process
depends on the details of the fluctuation spectrum. However, for quasi-coherent
chorus emission, we can simplify the first term on the right hand side of Eq. (\ref{eq:ballistic}) noting that 
the spectrum of $\omega_{\mathrm{tr} k}^{4} = k^2 v_\perp^2 (e \delta \bar B_{ky}/mc)^2$ is much narrower than
any characteristic widths due to $k = K(z=0,\omega)$ dependences in either group or phase velocities, as shown 
in Eq. (\ref{eq:chorusdisp}). Thus,
\begin{eqnarray*}
& & \sum_{k'} \frac{\left\langle\left\langle\omega_{\mathrm{tr} k'}^{4}\right\rangle\right\rangle}{4\left(1-v_{r \omega'} / v_{g \omega'}\right)^{4}}
\simeq \frac{\left\langle\left\langle (e k v_\perp /mc)^2 \right\rangle\right\rangle}{4\left(1-v_{r \omega} / v_{g \omega}\right)^{4}}\sum_{k'} \delta \bar B^2_{k'y} \\
& & \hspace*{2em} \simeq \frac{\left\langle\left\langle (e k v_\perp /mc)^2 \right\rangle\right\rangle}{4\left(1-v_{r \omega} / v_{g \omega}\right)^{4}}
\delta \bar B^2_{ky} = \frac{\left\langle\left\langle\omega_{\mathrm{tr} k}^{4}\right\rangle\right\rangle}{4\left(1-v_{r \omega} / v_{g \omega}\right)^{4}} \; .
\end{eqnarray*}
Here, we have noted that, from both observation and theory points of view, the magnetic fluctuation energy in a certain ``narrow'' $k$-spectrum range can be 
legitimately attributed to the reference $k$. Therefore, Eq. (\ref{eq:ballistic}) can be rewritten as 
\begin{equation}
\frac{\partial^{2}}{\partial t^{2}} \bar{\Gamma}_{N L}(\omega) \simeq\left(\frac{\partial \omega}{\partial t}\right)^2 \frac{\partial^{2}}{\partial \omega^{2}} \bar{\Gamma}_{N L}(\omega)
\; , \label{eq:ballistic2}
\end{equation}
where
\begin{equation}
\frac{\partial \omega}{\partial t}=\pm \frac{1}{2} \frac{\left\langle\left\langle\omega_{\mathrm{tr} k}^{4}\right\rangle\right\rangle^{1 / 2}}{\left(1-v_{r \omega} / v_{g \omega}\right)^{2}} \; ; \label{eq:choruschirp}
\end{equation}
and the corresponding ``self-similar'' solution is $\bar{\Gamma}_{N L}(\omega-(\partial \omega/\partial t) t)$, demonstrating 
its ballistic propagation in $\omega$-space at the rate given by Eq. (\ref{eq:choruschirp}).
This results is the analytic proof of the well-known Vomvoridis expression of chorus chirping \cite{Vomvoridis1982},
with $|R| = 1/2$ in Eq. (\ref{eq:R}) neglecting magnetic field non-uniformity,
based on the conjecture of maximized wave particle power transfer and PIC simulation results. 
Very similar chirping rates as in \cite{Vomvoridis1982} have also been obtained by \cite{Trakhtengerts2004,Omura2008}.
Equations (\ref{eq:ballistic}) and (\ref{eq:choruschirp}), originally derived in \cite{Zonca2021}, 
demonstrate that chorus chirping is indeed obtained by maximization of {\em spontaneous fluctuation growth};
\idest, maximization of wave particle power transfer \cite{Chen2016,Zonca2015b,Zonca2015}. They predict
that chorus can be rising- as well as falling-tone, consistent with \cite{Wu2020}; thus, suggesting that the
motivation of rising-tone chorus prevalence is due to the second term in the first square parenthesis on the right hand side
in Eqs. (\ref{eq:redWbar0}) and (\ref{eq:redGammabar0}), or other symmetry breaking effects (cf. next subsection).
Other possible explanations rely on the magnetic field non-uniformity \cite{Wu2020,Tao2021}, which is neglected
in the present simplified analysis. Interested readers are referred to original works for more details.
Another important feature of Eq. (\ref{eq:ballistic}) is that it shows that
the chorus chirping rate depends on the fluctuation spectrum and reduces to $|R| = 1/2$
only for a nearly monochromatic spectrum \cite{Zonca2021}. In other words, the actual
value of chorus chirping may depend on the initial hot electron distribution function
and other excitation conditions. Undoubtedly, for the conditions of Eq. (\ref{eq:ballistic}) to 
be derived from Eq. (\ref{eq:redGammabar0}), phase space structures must exist in the supra-thermal
electron phase space; that is, $|R| < 1$ is a stringent condition as discussed in Sec. \ref{sec:singlep}. 

\subsection{Reduced Dyson model for chorus emission: numerical solution}
\label{sec:reddyson_sol}

In this subsection, we numerically solve the reduced Dyson model equations 
schematically presented above, following the original derivation in \cite{Zonca2021}. For the sake of
convenience, we will treat the whistler wave fluctuation spectrum as a continuum, although
Eqs. (\ref{eq:redWbar0}) and (\ref{eq:redGammabar0}) have been derived assuming a 
generic discrete fluctuation spectrum, including a single mode as limiting case. 
For this, we introduce the normalized spectral density ${\cal I}(\omega)$ defined as \cite{Zonca2021}
\begin{equation}
\sum_{k} \frac{\left\langle\left\langle\omega_{t r k}^{4}\right\rangle\right\rangle}{\Omega_{e}^{4}} =
\sum_{k} \frac{\left\langle\left\langle k^3 v_\perp^2\right\rangle\right\rangle}{\Omega_{e}^{2}} \frac{I_k}{4 B_e^2}  \equiv \frac{\hat{\gamma}_{e}^{2}}{\Omega_{e}^{2}} \int \frac{d \omega}{\Omega_{e}} \mathcal{I}(\omega) \frac{\omega^{3 / 2}}{\left(\Omega_{e}-\omega\right)^{3 / 2}} \; , \label{eq:Inorm}
\end{equation} 
where $\hat \gamma_e$ is the peak value of the linear hot electron driving rate at the equator and normalizations
are chosen such that nonlinearity effects become important when ${\cal I}(\omega) \sim {\cal O}(1)$.
As in Sec. \ref{sec:reddyson_der}, only dependences on $\omega$ are explicitly given, while dependences on $t-z/v_{g\omega}$
are implicitly assumed for notation simplicity.
To practically solve the integral operators in the reduced Dyson model equations, let us introduce the auxiliary functions
\begin{eqnarray}
& &\left[ \frac{\left( \omega - \omega' \right)^2}{4 \Omega_e^2}+\Omega_e^{-2} \partial_t^2\right] G_{L1} (\omega, \omega') 
= \frac{\hat \gamma_e^2}{\Omega_e^2} \frac{{\cal I} (\omega)}{(1-v_{r\omega}/v_{g\omega})^4} \bar \Gamma_L (\omega) \nonumber \; , \\
& &\left[ \frac{\left( \omega - \omega' \right)^2}{4 \Omega_e^2}+\Omega_e^{-2} \partial_t^2\right] G_{L2} (\omega, \omega') 
= \frac{\hat \gamma_e^2}{\Omega_e^2} \frac{{\cal I} (\omega)}{(1-v_{r\omega}/v_{g\omega})^4} \bar \Gamma_L (\omega') \nonumber \; , \\
& &\left[ \frac{\left( \omega - \omega' \right)^2}{4 \Omega_e^2}+\Omega_e^{-2} \partial_t^2\right] G_{NL1} (\omega,\omega') 
= \frac{\hat \gamma_e^2}{\Omega_e^2} \frac{{\cal I} (\omega)}{(1-v_{r\omega}/v_{g\omega})^4} \bar \Gamma_{NL} (\omega) \nonumber \; , \\
& &\left[ \frac{\left( \omega - \omega' \right)^2}{4 \Omega_e^2}+\Omega_e^{-2} \partial_t^2\right] G_{NL2} (\omega,\omega') 
= \frac{\hat \gamma_e^2}{\Omega_e^2} \frac{{\cal I} (\omega)}{(1-v_{r\omega}/v_{g\omega})^4} \bar \Gamma_{NL} (\omega') \label{eq:Gdefs} \; ;
\end{eqnarray}
which let us rewrite Eq. (\ref{eq:redGammabar0}) and Eq. (\ref{eq:redWbar0}) respectively as
\begin{eqnarray}
\bar \Gamma_{NL} (\omega) & =  & \left[ \Omega_e \frac{\partial}{\partial \omega} - \frac{2 \Omega_e^2}{k^2 \left\langle v_\perp^2\right\rangle} 
\left( 1 - \frac{v_{r\omega}}{v_{g\omega}} \right) \right] \Omega_e \frac{\partial}{\partial \omega}  \int  \frac{d \omega'}{\Omega_e} \frac{\omega^{\prime 3/2}}{(\Omega_e - \omega')^{3/2}} \nonumber \\
& & \hspace*{-1em} \times  \frac{1}{8} \left( G_{L1} (\omega',\omega) + G_{L2} (\omega',\omega)
+ G_{NL1} (\omega',\omega) + G_{NL2} (\omega',\omega) \right)\label{eq:redGammabar2} ;
\end{eqnarray}
and
\begin{eqnarray}
\Omega_e^{-1} \partial_t \bar W (\omega) & =  & \left[ \Omega_e \frac{\partial}{\partial \omega} - \frac{2 \Omega_e^2}{k^2 \left\langle v_\perp^2\right\rangle} 
\left( 1 - \frac{v_{r\omega}}{v_{g\omega}} \right) \right] \Omega_e \frac{\partial}{\partial \omega}  \int  \frac{d \omega'}{\Omega_e} \frac{\omega^{\prime 3/2}}{(\Omega_e - \omega')^{3/2}}  \nonumber \\
& &  \hspace*{-5.5em} \times  \frac{\left( \omega' - \omega \right)}{16 \Omega_e} \left( G_{L1} (\omega',\omega) + G_{L2} (\omega',\omega)
+ G_{NL1} (\omega',\omega) + G_{NL2} (\omega',\omega) \right)\label{eq:redWbar1}  .
\end{eqnarray}
Recalling that we are focusing on the normalized spectral density evolution at fixed $z$ outside the hot electron source region,
Eqs. (\ref{eq:Gdefs}) to (\ref{eq:redWbar1}) can be closed by the intensity evolution equation,
\begin{eqnarray}
\Omega_e^{-1} \partial_t {\cal I} (\omega) & = & 2 {\cal S} {\cal I} (\omega)^{1/2} + 2 {\cal I} (\omega) \frac{\omega(\Omega_e-\omega)^2}{\Omega_e^3} \left[ \partial_t  \left( \int_{z-v_{g\omega}t}^\infty \zeta^4 (z') \frac{dz'}{v_{g\omega}} \right) \bar \Gamma(\omega) \right. \nonumber \\
& &\left.  \left.+ \left( \int_{z-v_{g\omega}t}^\infty \zeta^4 (z') \frac{dz'}{v_{g\omega}} \right) \partial_t \bar \Gamma (\omega) \right]\right/ \left( 1 - \frac{v_{r\omega}}{v_{g\omega}} \right)^2\; ; \label{eq:ievolve}
\end{eqnarray}
and the wave packet phase evolution equation, 
\begin{eqnarray}
\Omega_e^{-1} \partial_t \varphi (\omega) & = & - \frac{\omega(\Omega_e-\omega)^2}{\Omega_e^3} \left[ \partial_t  \left( \int_{z-v_{g\omega}t}^\infty \zeta^4 (z') \frac{dz'}{v_{g\omega}} \right) \bar W (\omega) \right. \nonumber \\
& &\left. \left.  + \left( \int_{z-v_{g\omega}t}^\infty \zeta^4 (z') \frac{dz'}{v_{g\omega}} \right) \partial_t \bar W (\omega) \right]\right/ \left( 1 - \frac{v_{r\omega}}{v_{g\omega}} \right)^2 \; ; \label{eq:phievolve}
\end{eqnarray}
which can be readily derived from Eqs. (\ref{eq:actionevolve}) and (\ref{eq:phaseevolve}) noting Eq. (\ref{eq:deryid1}) and the
analogous identity that one can write for $\partial_t \varphi$  \cite{Zonca2021}.
On the right hand side of Eq. (\ref{eq:ievolve}), we have added a source term $2 {\cal S} {\cal I} (\omega)^{1/2}$, representing 
the injection rate of fluctuations in the $|\delta \bar E_k|$ spectrum. In the absence of supra-thermal electrons, the effect of a constant source
would give an intensity spectrum ${\cal I} = {\cal S}^2 \Omega_e^2 t^2$, corresponding to $|\delta \bar E_k|$ linearly increasing with time. 
This choice gives us a variety of possibilities rather than assuming an initial spectrum.
Since Ref. \cite{Zonca2021} demonstrated that there are no qualitative differences between using ${\cal S}$ as a random source stirring the system or a constant uniform source in frequency, we focus, here, on the uniform source case.
\begin{figure}
	\begin{center}
		\resizebox{0.95\textwidth}{!}{\includegraphics{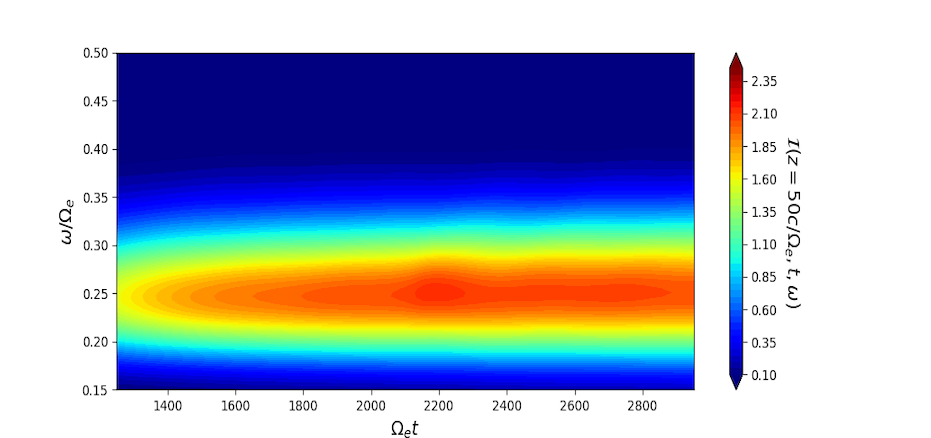}}
	\end{center}
	\vspace*{-2em}\noindent(a) 
	\begin{center}
		\resizebox{0.95\textwidth}{!}{\includegraphics{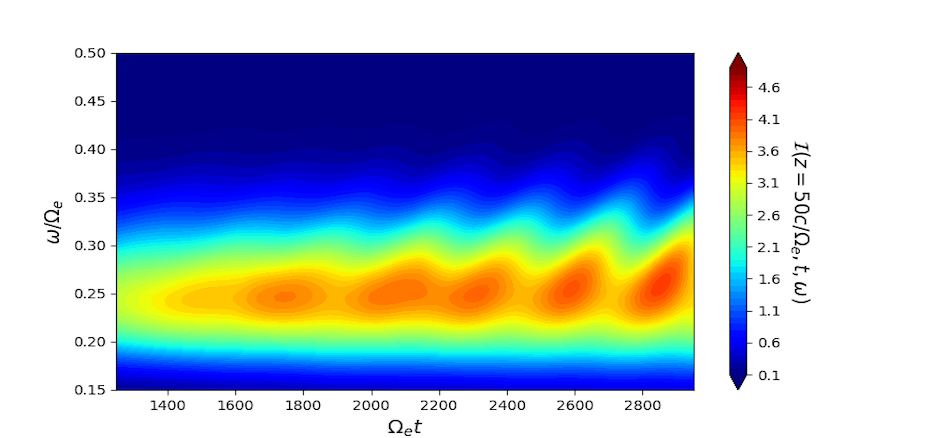}}
	\end{center}
	\vspace*{-2em}\noindent(b) 
	\begin{center}
		\resizebox{0.95\textwidth}{!}{\includegraphics{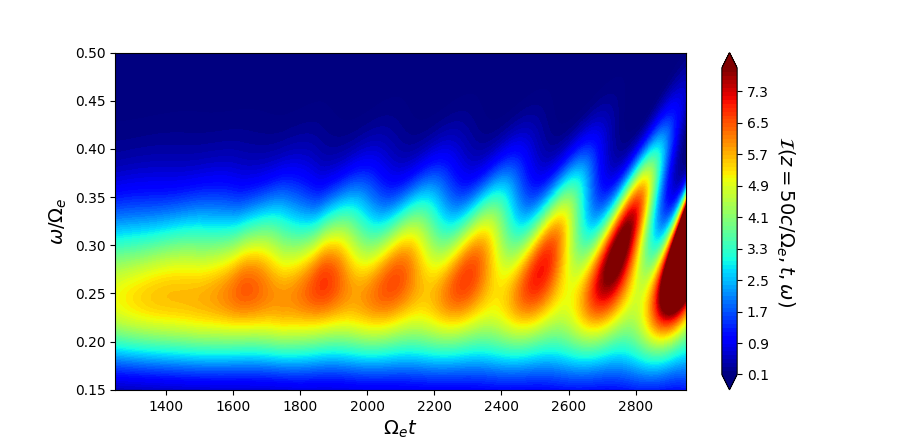}}
	\end{center}
	\vspace*{-2em}\noindent(c) 
\caption{Contour plots in the $(\Omega_e t, \omega/\Omega_e)$ space of the nonlinear evolution of the fluctuation spectrum for 
the same physical parameters as in Fig. \ref{fig:WGamma}. Position is fixed at 
$z \Omega_e/c = 50$, outside the hot electron source region. From top to bottom, 
the uniform (in $\omega$ space) source, ${\cal S}$ in Eq. (\ref{eq:ievolve}), increases: 
${\cal S} = 1 \times 10^{-5}$ [case (a)], ${\cal S} = \sqrt{2} \times 10^{-5}$ [case (b)], and
${\cal S} = 2 \times 10^{-5}$ [case (c)]. The existence of a lower threshold for the onset of chorus chirping
(cf. Sec. \ref{sec:chorusobs}) is evident when going from case (a) to case (b).}
\label{fig:NLspectrum}
\end{figure}

When solving Eqs. (\ref{eq:redGammabar2}) to (\ref{eq:phievolve}) by advancing them in time by
one step $\Delta t$, it is necessary to keep into account frequency ($\Delta_\omega = - \partial_t \varphi(\omega)$) and 
wave number ($\Delta_k$) shifts. Consistent with the analysis of Sec. \ref{sec:choruslin}, $\Delta_\omega/\Delta_k = v_{g\omega}$
and the nonlinearly shifted wave packet still satisfies the whistler dispersion relation. However, there exists
a corresponding small but finite shift in the resonant velocity $k \Delta v_{r\omega} = \Delta_\omega - v_{r\omega} 
\Delta_k = (1-v_{r\omega}/v_{g\omega})\Delta_\omega$. Thus, after advancing
by one step $\Delta t$, the functions $\bar \Gamma_{NL}$ and $\bar W_{NL}$ are actually evaluated at
$\omega + \Delta(\Delta_\omega)$, with $\Delta(\Delta_\omega) = - \Delta t \partial_t^2 \varphi(\omega)$, 
and, noting Eq. (\ref{eq:urel4}), these functions have to be updated as 
\begin{eqnarray}
\bar \Gamma_{NL} (\omega) & \rightarrow & \bar \Gamma_{NL} (\omega) - 
\Delta (\Delta_\omega) \frac{\partial}{\partial \omega} \bar \Gamma_{NL} (\omega) \; , \nonumber \\
\bar W_{NL} (\omega) & \rightarrow & \bar W_{NL} (\omega) - 
\Delta (\Delta_\omega) \frac{\partial}{\partial \omega} \bar W_{NL} (\omega) \; . \label{eq:WGammarenorm}
\end{eqnarray}
The same argument evidently applies to ${\cal I}(\omega)$ and $\varphi(\omega)$, such that
\begin{eqnarray}
{\cal I} (\omega) & \rightarrow & {\cal I} (\omega) - 
\Delta (\Delta_\omega) \frac{\partial}{\partial \omega} {\cal I} (\omega) \; , \nonumber \\
\varphi (\omega) & \rightarrow & \varphi (\omega) - 
\Delta (\Delta_\omega) \frac{\partial}{\partial \omega} \varphi (\omega) \; ; \label{eq:Iphirenorm}
\end{eqnarray}
after each time step. Equation (\ref{eq:Iphirenorm}) corresponds to changing $\partial_t \rightarrow \partial_t - \partial_t^2 \varphi(\omega) \partial_\omega$
on the left hand side of Eqs. (\ref{eq:ievolve}) and (\ref{eq:phievolve}); which corresponds to
solving the wave kinetic equation \cite{Bernstein1977,McDonald1988} for wave packet intensity and phase.
Initial conditions are homogeneous; that is, ${\cal I} (\omega) = 0$ together with  $\varphi (\omega) = 0$ at $t=0$,
as well as  $\left. G_{...}(\omega', \omega) \right|_{t=0} = 0$ and $\left. \partial_t G_{...}(\omega', \omega) \right|_{t=0} = 0$
on all auxiliary functions introduced in Eqs. (\ref{eq:Gdefs}). Meanwhile, solutions at the boundaries 
of the considered interval in $\omega$ space are obtained by linear
extrapolation of the inner solution. 

For the same physical parameters of Fig. \ref{fig:WGamma}, and assuming fixed $z \Omega_e/c = 50$, 
the nonlinear evolution of ${\cal I} (z = 50 c/\Omega_e, t, \omega)$ is shown in Fig. \ref{fig:NLspectrum} 
for ${\cal S} = 1 \times 10^{-5}$ [case (a)], ${\cal S} = \sqrt{2} \times 10^{-5}$ [case (b)], and
${\cal S} = 2 \times 10^{-5}$ [case (c)]. Here, we use a discretization in $\omega$ space with $221$ grid points in the interval 
$\omega/\Omega_e \in [0.05, 0.9]$ and adopt, for cases (a) and (b), a 
Savitzky--Golay filter fitting  sub-sets of 15 adjacent data points  with a fourth order degree polynomial to  
ensure regularity of the derivatives in $\omega$-space. Case (c) uses  fitting  sub-sets of 19 adjacent data points
to show that no significant changes are obtained by varying the Savitzky--Golay filter parameters. 
A fourth order Runge-Kutta integration in time is adopted
with variable time step, gradually decreasing from an initial $\Omega_e \Delta t = 1.25 \times 10^{-1}$ in the early linear evolution to 
$\Omega_e \Delta t = 3.125 \times 10^{-2}$ in the later nonlinear phase at $\Omega_e t > 1750$. This choice ensures that Courant condition
is well satisfied. The routine solving Eqs. (\ref{eq:Gdefs}) to (\ref{eq:phievolve}), closed by Eqs. (\ref{eq:WGammarenorm}) 
and (\ref{eq:Iphirenorm}) together with the aforementioned boundary 
conditions, is written in Python and uses Python standard libraries. 
Increasing ${\cal S}$ as in Fig. \ref{fig:NLspectrum} helps clarifying the role of driving strength for triggering
chorus emission; \idest, of the combination of convective amplification and initial intensity of the whistler wave
packet. A clear ``lower threshold'' is crossed from panel (a) to (b): in the former case, nonlinear oscillations
dominate the fluctuation spectrum evolution; while in the latter case, chorus elements are evidently produced. 
Figure \ref{fig:NLspectrum}, obtained by solution of the reduced Dyson model, is consistent with
the PIC simulation results of Fig. \ref{fig:whistler_cases} since, for the sake of simplicity, we have neglected 
resonance broadening and, thus, we do not expect chorus emission be smeared out by further increasing
the driving strength. 

\begin{figure}[t]
	\begin{minipage}{0.5\linewidth}
	\begin{center}
		\resizebox{\textwidth}{!}{\includegraphics{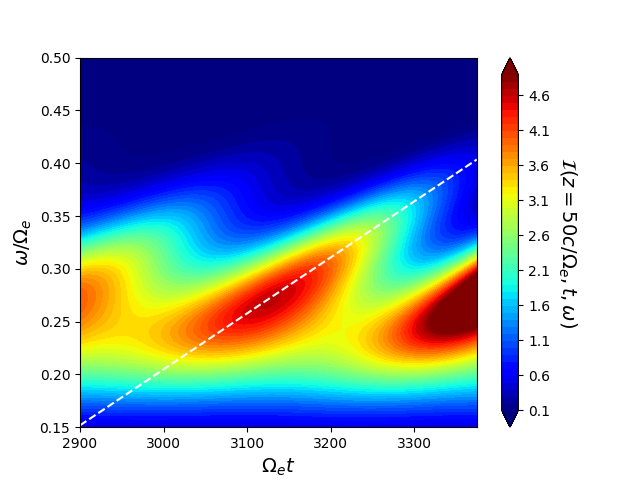}}
	\end{center} \end{minipage} \hfill \begin{minipage}{0.5\linewidth}
	\begin{center}
		\resizebox{\textwidth}{!}{\includegraphics{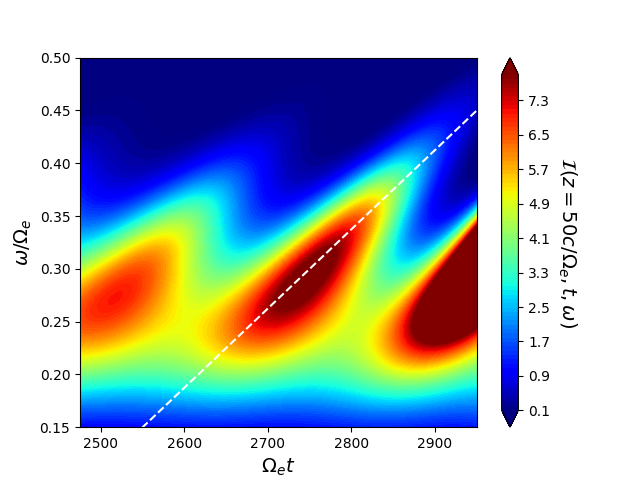}}
	\end{center} \end{minipage}
\vspace*{-2em}\newline\noindent(a) \hspace*{0.45\linewidth} (b)
\caption{Frequency chirping of individual chorus elements for ${\cal S} = \sqrt{2} \times 10^{-5}$ (a) and ${\cal S} = 2 \times 10^{-5}$ (b).
The white dashed lines passing
through the chorus elements represent the average chirping rate: $\partial_t \omega_0 = 5.3 \times 10^{-4}$ (a) and $\partial_t \omega_0 = 7.5 \times 10^{-4}$ (b).}
\label{fig:chirping}
\end{figure}

\begin{figure}[t]
	\begin{minipage}{0.5\linewidth}
	\begin{center}
		\resizebox{\textwidth}{!}{\includegraphics{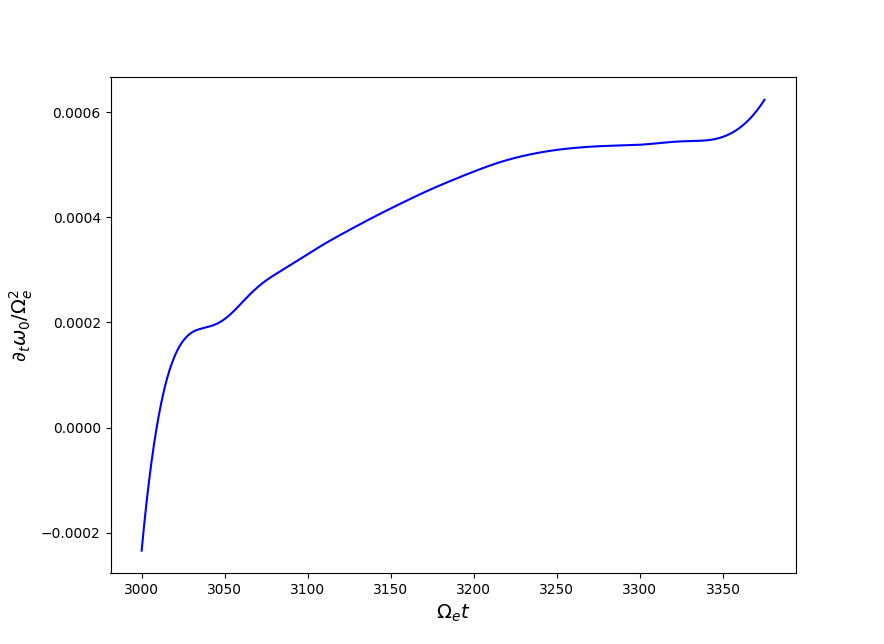}}
	\end{center} \end{minipage} \hfill \begin{minipage}{0.5\linewidth}
	\begin{center}
		\resizebox{\textwidth}{0.75\textwidth}{\includegraphics{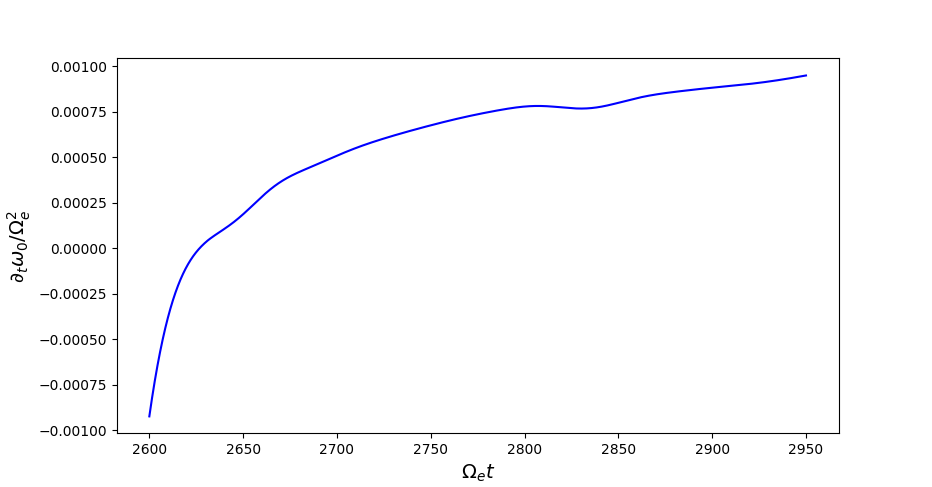}}
	\end{center} \end{minipage}
\vspace*{-2em}\newline\noindent(a) \hspace*{0.45\linewidth} (b)
\caption{Instantaneous chirping rate of the same chorus elements as in Fig. \ref{fig:chirping}: for ${\cal S} = \sqrt{2} \times 10^{-5}$ (a) and ${\cal S} = 2 \times 10^{-5}$ (b).}
\label{fig:chirprate}
\end{figure}

Focusing on frequency chirping, Fig. \ref{fig:chirping} shows: (a) the chorus element starting at $\Omega_e t \sim 3000$ 
for case ${\cal S} = \sqrt{2} \times 10^{-5}$; and (b) the chorus element starting at $\Omega_e t \sim 2600$ 
for case ${\cal S} = 2 \times 10^{-5}$.
The average chirping rate can be estimated as the average slope of a line fitting through the chorus element. It corresponds to the instantaneous chirping
rate becoming essentially constant after the intensity peak in the chorus element is reached. It 
is $\partial_t \omega_0 = 5.3 \times 10^{-4}$ for ${\cal S} = \sqrt{2} \times 10^{-5}$ and coincides 
with the value reported by PIC code simulations with DAWN code for the same parameters \cite{Tao2017a}. For ${\cal S} = 2 \times 10^{-5}$,
the average chirping rate $\partial_t \omega_0 = 7.5 \times 10^{-4}$, consistent with Eq. (\ref{eq:choruschirp}). The
behavior of the instantaneous chirping rate is reported in Fig. \ref{fig:chirprate}. The average values, 
$\partial_t \omega_0 = 5.3 \times 10^{-4}$ (a) and $\partial_t \omega_0 = 7.5 \times 10^{-4}$ (b), are clearly recognizable.
That instantaneous values of frequency chirping start from negative values in both cases is consistent with Eq. (\ref{eq:choruschirp}) and Ref. \cite{Wu2020};
and confirms that symmetry breaking that leads to the prevalence of rising tone chorus is due to the
second term in the first square parenthesis on the right hand side
in Eqs. (\ref{eq:redWbar0}) and (\ref{eq:redGammabar0}), as argued in Sec. \ref{sec:reddyson_der}, or other
effects such as the frequency asymmetry of the driving rate about its maximum, shown in Fig. \ref{fig:WGamma}(b).
Further reasons of symmetry breaking, connected with the non-uniformity of the background magnetic field are discussed
in \cite{Helliwell1967,Sudan1971,Wu2020,Tao2021} and are beyond the scope of the present analysis. We refer interested 
readers to the original works.

\begin{figure}[t]
	\begin{center}
		\resizebox{0.8\textwidth}{!}{\includegraphics{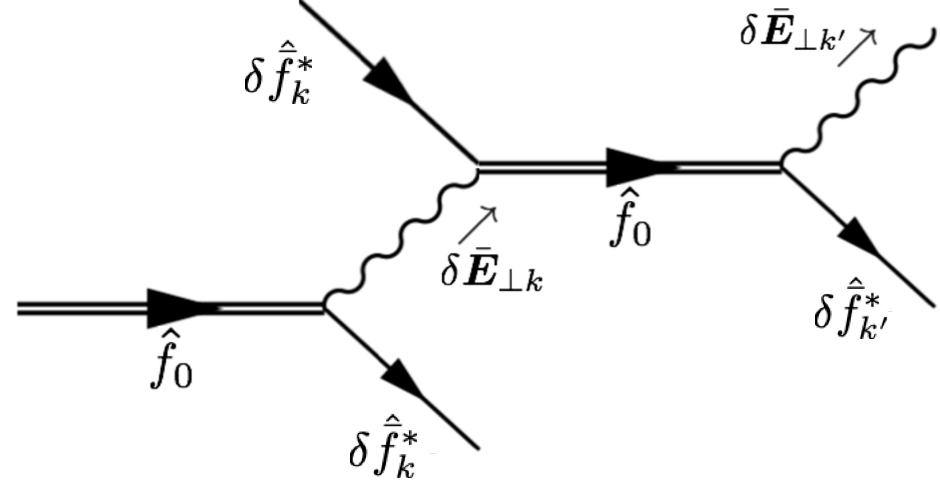}}
	\end{center}
\caption{Diagrammatic representation of chorus chirping consistent with Figs. \ref{fig:feynman} and \ref{fig:dyson} (adapted from the original 
figure from Ref. \cite{Zonca2021}). The double solid line propagator indicates
the hot electron renormalized response, which is unstable and {\em spontaneously emits} and reabsorbs most likely same-$k$ fluctuations corresponding to 
fluctuation spectrum instantaneous peak. Chorus chirping occurs because different $k$'s maximize wave particle power transfer at different times \cite{Zonca2017,Zonca2021}.}
\label{fig:dysonchirp}
\end{figure}

Nonlinear dynamics in one single chorus elements are due to maximization of wave particle power transfer and intensity growth,
as discussed in Sec. \ref{sec:reddyson_der} and expected for a {\em spontaneous process} \cite{Zonca2015b,Zonca2015,Chen2016,Zonca2021}.
Using the diagrammatic representation introduced in Figs. \ref{fig:feynman} and \ref{fig:dyson}, this process may be represented as
in Fig. \ref{fig:dysonchirp} \cite{Zonca2021}.
This is further demonstrated in Fig. \ref{fig:freshift}, where (a) provides the contour plot of the nonlinear frequency shift 
$\Delta_\omega(z=50c/\Omega_e,t)$, while (b) shows the snapshot of the same quantity at $\Omega_e t = 3125$. Here, 
we focus on the case at ${\cal S} = \sqrt{2} \times 10^{-5}$, since the detailed analysis of ${\cal S} = 2 \times 10^{-5}$ is given in \cite{Zonca2021}.
Figure \ref{fig:freshift} clearly shows that $\Delta_\omega$ is always much smaller than the dynamic range of frequency chirping, demonstrating
that individual oscillators in the whistler spectrum during a chorus event follow the well-know behavior of wheat explained by Leonardo Da Vinci:
\begin{figure}[t]
	\begin{minipage}{0.5\linewidth}
	\begin{center}
		\resizebox{\textwidth}{!}{\includegraphics{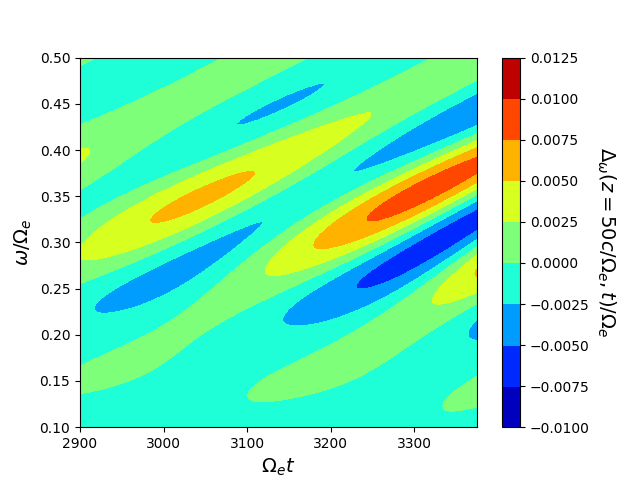}}
	\end{center} \end{minipage} \hfill \begin{minipage}{0.5\linewidth}
	\begin{center}
		\resizebox{\textwidth}{!}{\includegraphics{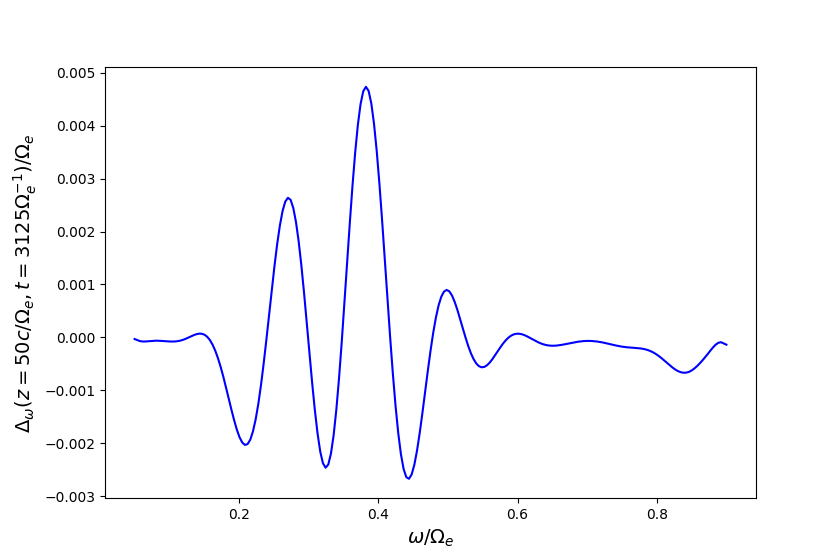}}
	\end{center} \end{minipage}
\vspace*{-2em}\newline\noindent(a) \hspace*{0.45\linewidth} (b)
\caption{Contour plot of the nonlinear frequency shift 
$\Delta_\omega(z=50c/\Omega_e,t)$ (a) and snapshot of the same quantity at $\Omega_e t = 3125$ (b) for the case ${\cal S} = \sqrt{2} \times 10^{-5}$
considered in Fig. \ref{fig:chirping}.}
\label{fig:freshift}
\end{figure}

\centerline{\em Accade sovente che l'onda si allontani dal suo punto di creazione,}
\centerline{\em mentre l'acqua non si muove, 
come le onde create dal vento } 
\centerline{\em in un campo di grano, dove vediamo le onde correre attraverso il campo}  
\centerline{\em mentre il grano rimane al suo posto}
\centerline{\em It often happens that waves travel away from their creation point,}
\centerline{\em while water does not move, like waves created by wind} 
\centerline{\em in a wheat field, where we see waves running across the field}  
\centerline{\em while wheat remains in place}

\begin{figure}[t]
	\begin{minipage}{0.5\linewidth}
	\begin{center}
		\resizebox{\textwidth}{!}{\includegraphics{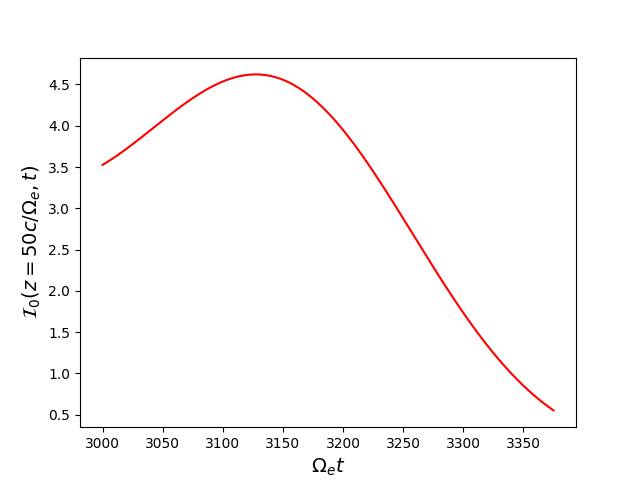}}
	\end{center} \end{minipage} \hfill \begin{minipage}{0.5\linewidth}
	\begin{center}
		\resizebox{\textwidth}{!}{\includegraphics{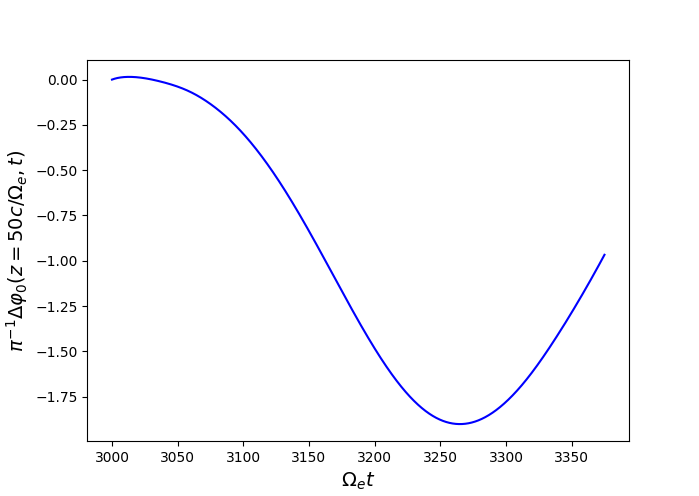}}
	\end{center} \end{minipage}
\vspace*{-2em}\newline\noindent(a) \hspace*{0.45\linewidth} (b)
\caption{Peak intensity, ${\cal I}_0(z=50c/\Omega_e,t)$ (a), and corresponding nonlinear phase shift $\Delta \varphi_0(z=50c/\Omega_e,t)$ (b), of the
chorus element starting at $\Omega_e t \sim 3000$ for the case ${\cal S} = \sqrt{2} \times 10^{-5}$
considered in Fig. \ref{fig:chirping}.}
\label{fig:element}
\end{figure}

\noindent The chorus element is the ``running wave'' and Fig. \ref{fig:element} illustrates the time behavior of its peak
intensity, ${\cal I}_0(z=50c/\Omega_e,t)$, and corresponding nonlinear phase shift, $\Delta \varphi_0(z=50c/\Omega_e,t)$, measured from the start of the considered
time interval. We will come back to this figure in Sec. \ref{sec:conclusions}. Here, it is worthwhile noting that the maximum 
of the intensity is reached when $\Delta \varphi_0\simeq - \pi$ and that the nonlinear time change in fluctuation intensity
is also reflected by the similar change in space, due to the dependence of ${\cal I}$ on $t - z/v_{g\omega}$.
The same behavior is obtained for ${\cal S} = 2 \times 10^{-5}$ \cite{Zonca2021}.

\subsection{Comparison with other models of chorus chirping}
\label{sec:compare}

The comparison with the analysis of chorus chirping originally proposed by \cite{Vomvoridis1982}
has been extensively provided in the previous subsections. One of the merits of the present theoretical
framework is the analytical derivation, Eq. (\ref{eq:choruschirp}), of the optimal condition for wave particle 
power transfer, consistent with the conjecture of Ref. \cite{Vomvoridis1982} based on PIC simulation 
results. We have also briefly analyzed the role of background magnetic field non-uniformity, originally
discussed by \cite{Helliwell1967,Sudan1971} and more recently revisited by \cite{Wu2020,Tao2021}.
What may seem in apparent contrast with the present theoretical analysis is the interpretation of 
chorus frequency chirping given by \cite{Omura2011}, explained as consequence of  
the nonlinear current parallel to the wave magnetic field ($J_B$), responsible of 
nonlinear frequency shift. In particular, a sequence of ``whistler seeds'' that are excited and 
amplified by wave particle resonant interactions are at origin
of frequency chirping. 

\begin{figure}[t]
	\begin{minipage}{0.5\textwidth}
		\centerline{\resizebox{\textwidth}{!}{\includegraphics{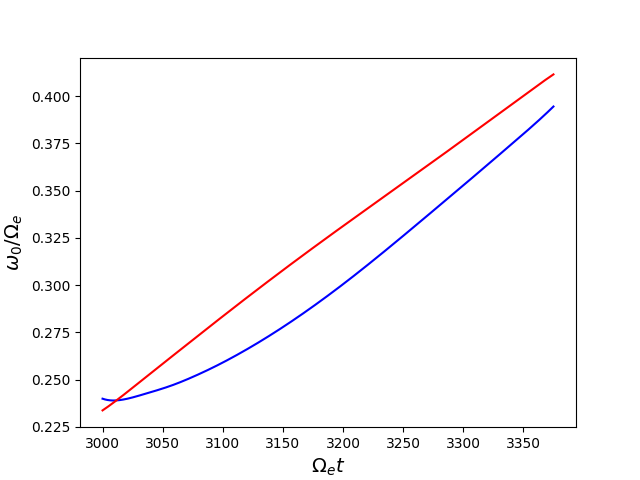}}}
	\end{minipage} \hfill \begin{minipage}{0.5\textwidth}
		\centerline{\resizebox{\textwidth}{0.75\textwidth}{\includegraphics{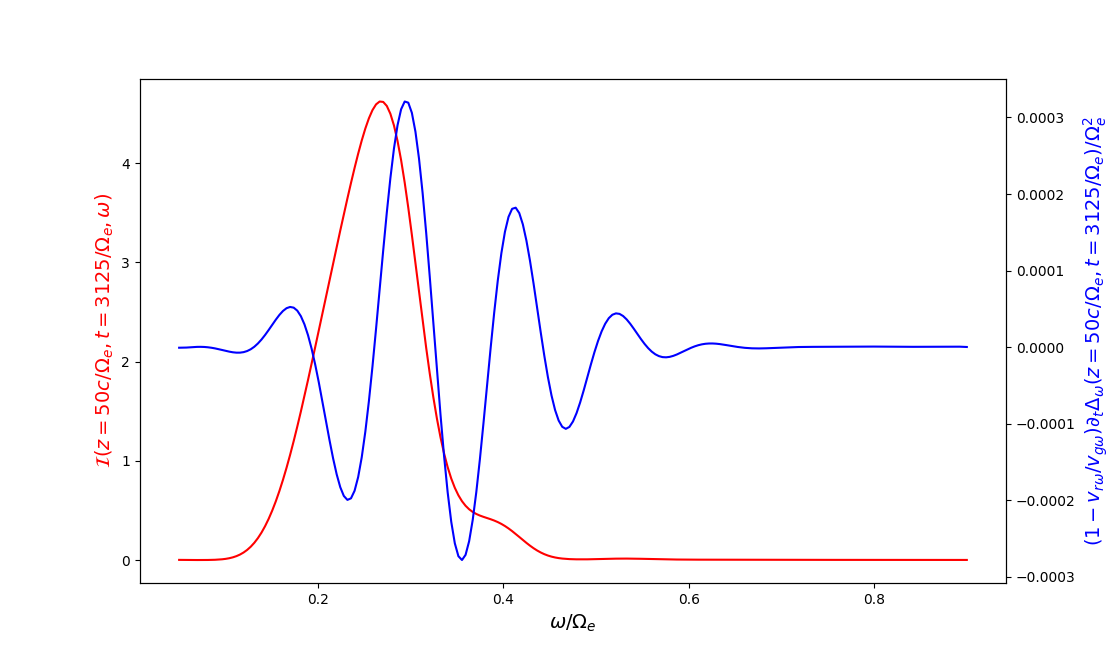}}}
	\end{minipage}
	\vspace*{-.5em} (a) \hspace*{0.46\textwidth} (b)
\caption{(a) Time evolution of the intensity peak frequency (blue line) of the chorus element analyzed in Fig. \ref{fig:element}.
The red line denotes the frequency evolution of the maximum in the rate of change of nonlinear frequency shift. (b) 
Snapshot at $\Omega_e t =3125$ of the fluctuation intensity and of the $\partial_t \omega_{\rm res} = (1-v_{r\omega}/v_{g\omega})
\partial_t \Delta_\omega$ as a function 
of frequency.} 
\label{fig:chirprel}
\end{figure}

Within the present theoretical framework, the fluctuation spectrum is self-consistently evolved 
and Fig. \ref{fig:freshift} demonstrates that each oscillator in the wave spectrum (each ``wheat plant''
in Leonardo's metaphor; cf. Sec. \ref{sec:reddyson_sol}) is characterized by a small nonlinear frequency shift.
However, continuing the metaphor, when the ``running wave'' crosses the field while wheat remains in place, each
wheat plant swings in tune with the wave when it is passed by the fluctuation peak. This also occurs for the
chorus elements and it is shown in Fig. \ref{fig:chirprel}(a), where 
the time evolution of the intensity peak frequency of the chorus element analyzed in Fig. \ref{fig:element} is represented by the blue line.
The red line, meanwhile denotes the frequency evolution of the maximum in the rate of change of nonlinear frequency shift.
The connection between the frequencies of these two peaks is also shown in Fig. \ref{fig:chirprel}(b), providing the 
snapshot at $\Omega_e t =3125$ of the fluctuation intensity and of $\partial_t \omega_{\rm res} = (1-v_{r\omega}/v_{g\omega})
\partial_t \Delta_\omega$ as a function of frequency. Recalling the discussion preceding Eq. (\ref{eq:WGammarenorm}), $\partial_t \omega_{\rm res}$
represents the rate of change of the resonance frequency.
In conclusion, if one interprets the ``whistler seeds'' of \cite{Omura2011} as the individual oscillators 
in the wave packet at the intensity peak, one should obtain the frequency increase due to the chorus chirping as 
$$ \Delta \omega = \int  (1-v_{r\omega_0(t')}/v_{g\omega_0(t')})\partial_{t'} \Delta_{\omega_0(t')} dt' \; , $$
where integration is to be intended along the red line of Fig. \ref{fig:chirprel} (a) \cite{Zonca2021}.
The hence obtained frequency increase is $\Delta \omega/\Omega_e = 0.121$ over the considered 
time interval, against the corresponding frequency shift, $\Delta \omega/\Omega_e = 0.155$,  of the intensity peak.
We believe that this good agreement confirms the present explanation that reconciles the original interpretation of 
frequency chirping given by \cite{Omura2011} with the present theoretical analysis.

As a final remark, we would like to briefly comment about the recent work by
\cite{Tsurutani2020}, discussing Van Allen Probe data and showing  that each chorus element is made
of discrete sub-elements with constant frequency. The present theoretical framework, and  Fig \ref{fig:freshift} in particular, 
supports that each oscillator in the whistler wave spectrum has a nonlinear frequency shift in the order of a few percent, 
consistent with observations by \cite{Tsurutani2020}. 
The discrete steps, which are the essential elements of the rising tone chorus element in this recent work, are instead beyond the
description of the present theoretical analysis of the reduced Dyson model equations since, by definition in Eq. (\ref{eq:Inorm})
and for the sake of simplicity, 
we assume the continuous limit. However, adopting the general theoretical framework discussed in Sec. \ref{sec:phasespace},
it would be possible to solve the self-consistent evolution equations for the fluctuation spectrum in any arbitrary discretized form.
This approach would, thus, allow us to address the conditions described in
\cite{Tsurutani2020}. Doing so is, however, beyond the scope of the tutorial style of the present work.

For the sake of completeness, we also mention the recent and extensive analysis by \cite{Zhang2020a,Zhang2020b}, which provides
statistical results from 6 years of Van Allen Probes data on the value of $\partial \omega/\partial t$ inside chorus wave packets.
Unlike Ref. \cite{Tsurutani2020},  \cite{Zhang2020a,Zhang2020b} show that the frequency variation inside long packets ($> 50$ wave periods) is generally non-null and consistent with the usual chirping rate of \cite{Vomvoridis1982,Omura2008} and the present Eq. (\ref{eq:choruschirp}). These statistical 
observations have been recovered in numerical simulation results by \cite{Nunn2021}. Meanwhile, for very short packets ($< 10$ wave periods),
the results by \cite{Zhang2020a,Zhang2020b} agree with earlier works that attribute frequency chirping to nonlinear wave-particle interactions
and wave superposition for moderate amplitudes \cite{Katoh2016,Crabtree2017}.

\section{Discussion and conclusions}
\label{sec:conclusions}

The physics processes connected with chorus emission, discussed in the previous sections,
are so deeply rooted in fundamental elements of nonlinear dynamics and phase space transport 
that one may naturally expect many points of contact with similar phenomena occurring
in space and laboratory plasmas. Following the intended scope of the present work, we
will concentrate on the latter, discussing briefly examples from {\em free electron lasers} (FEL) and
{\em magnetic confinement fusion} (MCF). 
Readers who may be interested in a somewhat ``unconventional'' perspective on fusion, triggered by collapsing
whistler waves, are referred to the recent work by \cite{Sano2020}.

Analogies and similarities of chorus emission and FEL physics was noted and addressed in Ref.
\cite{SotoChavez2012}. Here, we follow Refs. \cite{Zonca2015b,Zonca2015,Chen2016} and the
more recent Ref. \cite{Tao2021,Zonca2021} and draw analogies with FEL {\em super-radiance} 
\cite{bonifacio90,bonifacio94,giannessi05,watanabe07}.
The dynamic evolution of intensity peak and phase of the chorus element in Fig. \ref{fig:element}
correspond to synchronization (cf. Ref. \cite{escande18}) of supra-thermal electrons and 
subsequent detuning as the chorus wave packet slips over the resonant electron population. During
synchronization, phase evolution is minimized so that wave particle power extraction is 
maximum. While resonant particles release their energy to the wave, by moving from 
the positive to negative $\dot \phi$ domain in Fig. \ref{fig:wavetrap} as discussed in Sec. 
\ref{sec:chorusobs}, peak intensity grows and phase gradually reaches $-\pi$. 
Resonance detuning, then, becomes relevant and is made more significant by
the drop in fluctuation intensity. The analogies between these phases taking place during
a chorus element and nonlinear pulse evolution in seeded FEL amplifiers, respectively, are further
clarified by direct comparison of simulation results discussed in Fig. 6 of Ref. \cite{Tao2021}
and Fig. 3 of Ref. \cite{giannessi05}. The main and most significant difference between
chorus emission in planetary magnetospheres and FEL super-radiance is that the former
is a {\em spontaneous} process, while the latter is {\em stimulated} or externally controlled.
Interested readers are referred to original works for further details. 
In addition to this, it is worthwhile mentioning that parametric whistler wave generation during laser-plasma interaction 
have been addressed by \cite{Mourenas1998}.
We also note that the theoretical approaches 
presented in this work may also be applied toward in-depth understandings 
of the nonlinear phenomena in high power radiation devices such as 
gyrotron backwave oscillators \cite{SChen2012,SChen2013}.

The strong connection of physics processes underlying chorus emission and various
phenomena observed in MCF is actually one of the primary motivations of the present work,
as mentioned in the Introduction. Magnetized plasmas of fusion interest are, in fact, the 
natural environment where a number of {\em spontaneous} frequency sweeping phenomena
occur, since fluctuations of the shear Alfv\'en wave spectrum are strongly excited by 
resonant supra-thermal particles, notably charged fusion products and/or
particles accelerated by  additional heating systems \cite{Zonca2015b,Zonca2015,Chen2016}. 
For low-frequency fluctuations (below the ion-cyclotron frequency),
self-consistent single-particle nonlinear dynamics in MCF plasmas reduces to 
a non-autonomous system with two degrees of freedom, while the same problem
in the chorus case is well described by a non-autonomous system with one degree
of freedom. In addition to this, the complex toroidal geometry and the nonlocal 
plasma response in MCF require analyzing the fluctuation spectrum evolution 
by means of a nonlinear Schr\"odinger-like equation (NLSE) \cite{Zonca2015b,Zonca2015,Chen2016},
well beyond the wave kinetic equation paradigm \cite{zonca21}. The systematic development for MCF
of the same theoretical framework presented in Sec. \ref{sec:phasespace} has been carried out
by \cite{Zonca2015b,Zonca2015,Chen2016} and, more recently, Ref. \cite{falessi19a}, 
leading to the so-called {\em Dyson-Schr\"odinger Model} (DSM) for
transport in MCF \cite{zonca21}. The aforementioned complexities due to number of degrees
of freedom and equilibrium geometry require a more systematic numerical approach
to the solution of general governing equations. Notwithstanding these complications,
solutions of simple limits in cases of practical interest have been obtained, such as 
for Energetic Particle Modes (EPM) \cite{Zonca2015} and fishbones \cite{Chen2016}. In
these cases, in particular,
the existence of self-similar structures characterized by ballistic propagation in phase space 
establishes a fundamental connection between chorus and EPM as well as fishbone frequency
chirping in MCF \cite{Chen2016,Zonca2015}. In fact, the predicted propagation speed of these resonant structures
scales linearly with the phase space velocity (and fluctuation amplitude)
in the ``resonant action'' direction; that is, velocity (and/or frequency)
for chorus and canonical angular momentum for EPM/fishbone.
Interested readers may find a recent and in depth review of these analyses in 
Ref. \cite{Chen2016}. 

As a final remark of this short trip through chorus physics, taken aboard the comfortable
environment of plasma theory, we would like to note that windows were opened
with a view over existing problems of both academic interest and practical implications.
The mathematical language in which these problems are formulated are the Dyson
Schwinger Equation (DSE) (cf. \exgra \ \cite{Itzykson80}) and NLSE. This naturally opens
points of contact with condensed matter physics and more. The peculiar feature of
collisionless plasmas to be characterized by nonlocal response in space and time
gives to nonlinearities in these equations an intrinsic integral nature and is cause 
of a practically infinite variety of interesting behaviors. This is ultimately what makes
plasma physics such an exciting cross-disciplinary field to study.

\section*{Acknowledgements}
Useful discussions with D. F. Escande are kindly acknowledged. 
This work was carried out within the framework of the EUROfusion Consortium 
and received funding from Euratom research and training programme 2014--2018 
and 2019--2020 under Grant Agreement No. 633053 (Project No. WP19-ER/ENEA-05). 
The views and opinions expressed herein do not necessarily reflect those of the European Commission.
This work was also supported by NSFC grants (41631071 and 11235009), the Strategic Priority Program 
of the Chinese Academy of Sciences (No. XDB41000000)
and the Fundamental Research Funds for the Central Universities.
% \end{acknowledgements}

%\appendix

%\section{Appendix: Equations for strongly excited convective cells}

\section*{Conflict of interest}

On behalf of all authors, the corresponding author states that there is no conflict of interest.

\bibliographystyle{spbasic}      % basic style, author-year citations
\bibliography{rmppbib}   % name your BibTeX data base

\end{document}